\documentclass[a4paper,fleqn,usenatbib]{mnras}

\usepackage{newtxtext,newtxmath}

\usepackage[T1]{fontenc}
\usepackage{ae,aecompl}

\usepackage{graphicx}	
\usepackage{amsmath}	
\usepackage{amssymb}	
\usepackage{csquotes}

\title[Stellar Mass Assembly of Low-z, Massive Centrals]{The Stellar Mass Assembly of Low Redshift, Massive, Central Galaxies in SDSS and the TNG300 simulation}

\author[Thomas M. Jackson et al.]{
	Thomas M. Jackson$^{1}$\thanks{E-mail: Thomas.jackson@uni-heidelberg.de},
	A. Pasquali$^{1}$,
	C. Pacifici$^{2}$,
	C. Engler$^{1, 3}$,
	A. Pillepich$^{3}$, 
	\newauthor{E. K. Grebel$^{1}$}
	\\
	$^{1}$Astronomisches Rechen-Institut, Zentrum f{\"u}r Astronomie der Universit{\"a}t Heidelberg, M{\"o}nchhofstr. 12-14, D-69120 Heidelberg, Germany\\
	$^{2}$Space Telescope Science Institute, 3700 San Martin Drive, Baltimore MD 21218, USA\\
	$^{3}$Max-Planck-Institut f{\"u}r Astronomie, K{\"o}nigstuhl 17, D-69117 Heidelberg, Germany
}

\date{Accepted 2020 July 31. Received 2020 July 31; in original form 2020 February 25.}

\pubyear{2020}

\begin{document}
	\label{firstpage}
	\pagerange{\pageref{firstpage}--\pageref{lastpage}}
	\maketitle
	
	\begin{abstract}
		The stellar mass assembly of galaxies can be affected by both secular and environmental processes. In this study, for the first time, we investigate the stellar mass assembly of $\sim90,000$ low redshift, central galaxies selected from SDSS group catalogues (M$_{\rm Stellar}\gtrsim10^{9.5}$M$_{\odot}$, M$_{\rm Halo}\gtrsim10^{12}$M$_{\odot}$) as a function of both stellar and halo mass. We use estimates of the times at which 10, 50 and 90 per cent of the stellar mass was assembled from photometric spectral energy distribution fitting, allowing a more complete investigation than single stellar ages alone. We consider trends in both stellar and halo mass simultaneously, finding dependencies of all assembly times on both. We find that galaxies with higher stellar masses (at constant halo mass) have on average older lookback times, similar to previous studies of galaxy assembly. We also find that galaxies at higher halo mass (at constant stellar mass) have younger lookback times, possibly due to a larger reservoir of gas for star formation. An exception to this is a sub sample with high stellar-to-halo mass ratios, which are likely massive, field spirals. We compare these observed trends to those predicted by the TNG300 simulation, finding good agreement overall as a function of either stellar or halo mass. However, some differences in the assembly times (of up to $\sim 3$ Gyr) appear when considering both stellar and halo mass simultaneously, noticeably at intermediate stellar masses (M$_{\rm Stellar} \sim 10^{11}$ M$_{\odot}$). These discrepancies are possibly linked to the quenched fraction of galaxies and the kinetic mode AGN feedback implemented in TNG300. 
		
	\end{abstract}
	
	
	
	
	\section{Introduction}
	\label{sec:intro}
	
	Studies of massive galaxies have found that despite their seemingly simple structure, they have more complex assembly histories. Observations claim a short, quick burst of star formation at high redshift \citep{2005ApJ...621..673T, 2006ARA&A..44..141R} during which the bulk of the stellar mass is formed. Observations also reveal that massive high redshift galaxies appear to have smaller physical sizes than their low redshift counterparts \citep{2005ApJ...626..680D, 2006MNRAS.373L..36T}, meaning that this population of galaxies appears to significantly increase in size while gaining relatively little mass from early epochs \citep{2010ApJ...709.1018V, 2014ApJ...788...28V}. 
	
	Theories of galaxy formation have been used to link these different observational results \citep{2009ApJ...699L.178N}. They suggest a two-phase scenario, whereby massive galaxies in the early universe rapidly form the bulk of their stellar mass through a main dissapative episode of star formation and later increase their physical size via minor mergers \citep[e.g.][]{2009ApJ...697.1290B}, indicating that both secular and environmental processes play a significant role in galaxy formation and evolution.
	
	A technique commonly used to constrain the mass assembly processes of galaxies is the use of stellar population synthesis applied to spectroscopic data. These techniques employ a stellar library, either empirical \citep{2010MNRAS.404.1639V} or theoretical \citep{2003MNRAS.344.1000B, 2005MNRAS.362..799M}, in order to synthesise a combination of single-age stellar spectra. These combinations are then matched to the observed spectrum in order to find the weighted best-fitted age estimation, among other characteristics such as metallicity or [$\alpha$/Fe], which can reveal more about a galaxy's assembly history \citep[see][for an overview]{2009ApJ...699..486C}. 
	
	Studies utilising this technique have found evidence that more massive galaxies form their stars earlier and on shorter timescales than less massive ones \citep{2005ApJ...621..673T, 2005MNRAS.362...41G, 2010MNRAS.404.1775T}, commonly known as downsizing \citep{1996AJ....112..839C}. These techniques have also been applied to both long slit spectroscopy, to investigate radial gradients of age, metallicity, IMF etc. \citep{2012MNRAS.426.2300L, 2019MNRAS.489.4090L, 2015MNRAS.447.1033M} and more recently to integral field spectroscopy \citep{2015MNRAS.449..328W, 2020MNRAS.491.3562Z} to explore galaxy assembly and evolution processes in more detail. There are, however, some restrictions associated with these techniques. The sample sizes can be significantly smaller compared to photometric studies of galaxy evolution. They may also focus mainly on early-type galaxies rather than the general galaxy population due to the problems in breaking degeneracies in dust, age and metallicity that dominate late-type galaxies as well as the generally more complicated star formation histories which cannot be well described by a single stellar population \citep[see][and references therein]{2009ApJ...699..486C}. 
	
	An alternative technique is to use photometric spectral energy distribution fitting \citep[hereafter SED fitting, see e.g.][]{2008MNRAS.388.1595D}. \citet{2012MNRAS.421.2002P} used this technique, implementing it on a large sample of galaxies from SDSS \citep{2016ApJ...824...45P}. Their method uses semi-analytic models to generate a wide range of star formation histories and fit these to observed photometry. This allows estimations of the times at which certain percentages of stellar mass were assembled, exploring in more detail the assembly histories of certain galaxies, although with larger uncertainties than spectroscopic techniques. The advantage of this technique is that photometric catalogues are generally significantly larger than spectroscopic catalogues, allowing for a greater range of applications.
	
	Estimated ages or lookback times are then usually compared as a function of stellar mass, which is postulated to be the main driver in galaxy evolution, especially in central galaxies
	\citep{2005MNRAS.362...41G, 2010MNRAS.407..937P}. Multiple processes are driven predominantly by stellar mass, ranging from star formation \citep{1998ApJ...498..541K} to the quenching of star formation via supernova or Active Galactic Nuclei (AGN) feedback \citep[see e.g.][]{2006MNRAS.370..645B}. Environment, however, is also known to have an effect on galaxy assembly \citep{2010MNRAS.407..937P}. These processes range from major and minor mergers supplying molecular gas, resulting in starbursts \citep[e.g.][]{1996ARA&A..34..749S}, to processes which shut down star formation such as strangulation and ram-pressure stripping \citep[see e.g.][]{2011ASSP...27..203D}.
	
	This variety of processes can affect many of the same galaxy properties meaning there may be many degeneracies between the two drivers of galaxy evolution. By comparing galaxy properties as a function of both simultaneously, we can attempt to somewhat separate secular evolution from environmental effects. To do this, two proxies for the environmental and secular processes are commonly used, namely the stellar mass of the galaxy itself for secular processes and the parent halo mass of a galaxy for the environment \citep{2006MNRAS.366....2W}. Galaxy properties are then averaged in bins of stellar and halo mass, or compared in a number of different fixed stellar (halo) masses while varying halo (stellar) mass. This concept has been employed in previous studies in order to investigate galaxy morphological types \citep{2006MNRAS.366....2W}, activity types \citep{2009MNRAS.394...38P}, star formation rates \citep{2008arXiv0805.0002V, 2012MNRAS.424..232W}, ages and metallicities \citep{2014MNRAS.445.1977L, 2020arXiv200601154T} and satellite assembly and evolution \citep{2019MNRAS.484.1702P, 2019ApJ...876..145S} and their dependencies on both halo and stellar mass.
	
	Comparing observations to simulations is advantageous in such studies for a number of reasons: Simulations can track the entire stellar material of galaxies throughout cosmic time, allowing a full construction of the star formation and mass assembly histories \citep{2007MNRAS.375....2D}. Therefore, by comparing trends in both data sets, we can improve the modelling within simulations, leading to more accurate predictions. We can also use simulations to better constrain and estimate the physical processes driving the evolution of galaxies. However, when employing these sorts of techniques, a deep level of matching is needed and a realisation of the limits for both observations and simulations needs to be clearly acknowledged in order to avoid over-interpreting results. 
	
	In this paper we present the stellar mass build-up within approximately 2 effective radii ($\sim$ 2$R_e$) of central galaxies. Our observational sample contains $\sim90,000$ low redshift ($z < 0.15$), central galaxies selected from SDSS group catalogues of \citet{2017MNRAS.470.2982L}, \citet{2007ApJ...671..153Y} and \citet{2005AJ....129.2562B}. From these catalogues we obtain both environmental and host galaxy properties. We then use multiple stellar mass assembly times, estimated via SED fitted models from the work of \citet{2016ApJ...824...45P}. This allows us to investigate for the first time how characteristic assembly histories (instead of single stellar ages) correlate with both environmental and secular factors. Finally, we compare the findings inferred from the observational data with the results of the TNG300 simulation \citep{2018MNRAS.475..648P, 2018MNRAS.480.5113M, 2018MNRAS.475..676S, 2018MNRAS.475..624N, 2018MNRAS.477.1206N} in order to see how well the simulations reproduce what we observe and what processes may shape the observational trends. 
	
	In Section~\ref{sec:Data} we present the observational data and methods used in this research. In Section~\ref{sec:Simulations} we briefly introduce the simulation used in this research and the criteria used to select the comparison data. We then present our results and comparisons in Section~\ref{sec:Results}. In Section~\ref{sec:Discussion} we discuss these results before summarising our work in Section~\ref{sec:conclusions}.

	\section{Observational data and methods}
	\label{sec:Data}
	
	\subsection{SDSS group catalogues}
	\label{data:catalogues}
	
	Group catalogues are extremely useful for exploring the characteristic properties of various sub-populations of galaxies. This is due to their statistically significant sample sizes, which reduce biases introduced by outliers and reveal an overall picture of galaxy evolution and its possible dependencies on both secular and/or environmental processes. In this research we used the SDSS group catalogues of \citet{2017MNRAS.470.2982L} which build upon the group catalogues of \citet{2007ApJ...671..153Y}. We give a brief outline of the catalogues here: A full description of the entire catalogues and techniques can be found in \citet{2017MNRAS.470.2982L}. 
	
	\begin{figure}
		\centering
		\includegraphics[width=1.1\columnwidth]{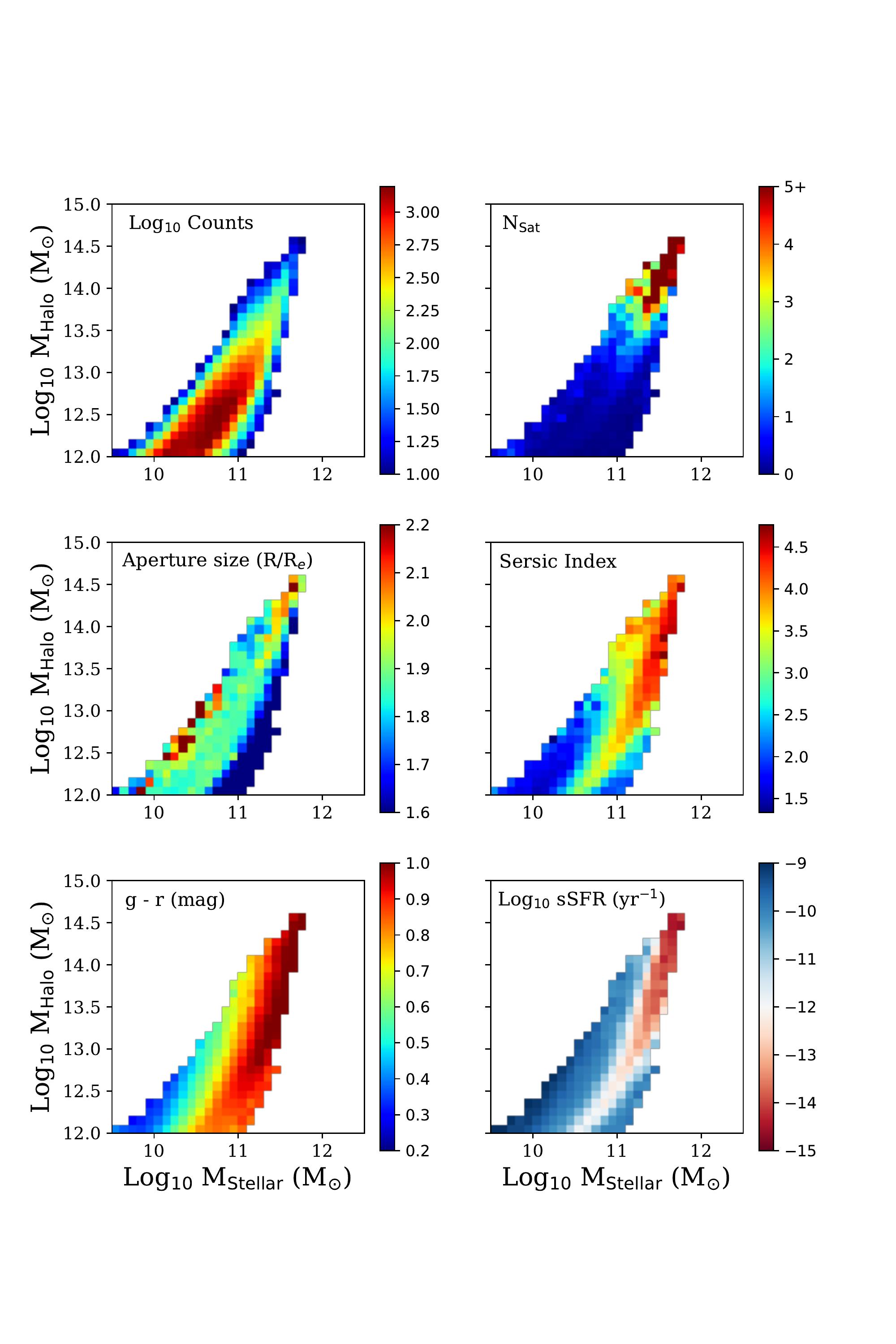}
		\caption{The average properties of our SDSS galaxy sample per bin in both stellar and halo mass. From top left to bottom right we show the logarithm of the counts of galaxies, the average number of satellites (N$_{\mathrm{Sat}}$), the ratio of the radius in which the photometry was measured to the effective radius, the average S\'{e}rsic index, the mean $g - r$ colour and the sSFR estimated from the SED fitting.}
		\label{fig:properties}
	\end{figure}
	
	The catalogues of \citet{2017MNRAS.470.2982L} use data from the SDSS data release 13 \citep[DR13,][]{2017ApJS..233...25A}, which improves on the data release 7 \citep[DR7,][]{2009ApJS..182..543A} used in \citet{2007ApJ...671..153Y}. Two of the main improvements are in the photometry, which implements refined reduction of the raw data and uses updated zero-points, and the multi-object spectroscopy, which obtains spectroscopic redshifts for some objects that were missed in DR7 due to fibre collisions. \citet{2017MNRAS.470.2982L} also use an improved Friends-of-Friends (FoF) algorithm to link galaxies to their parent haloes. This has been compared to \citet{2007ApJ...671..153Y}, with no significant differences in the majority of groups, however better constraints on the halo masses of groups with low numbers of galaxies. 
	
	Each galaxy is initially treated individually while the algorithm estimates halo masses based on the stellar masses. These estimates are generated in mock catalogues from the EAGLE simulation \citep{2015MNRAS.446..521S, 2015MNRAS.450.1937C}, which assumes a $\Lambda$CDM cosmology and returns dark matter halo profiles consistent with \citet*[][NFW]{1997ApJ...490..493N} profiles. Line-of-sight velocity dispersions ($\sigma_{\mathrm{LoS}}$) and halo radii ($r_{180}$, the radius of the halo where the mean mass density is 180 times the mean density of the Universe) are then estimated. A probability function of two galaxies belonging to the same parent halo is derived using $\sigma_{\mathrm{LoS}}$ and $r_{180}$ and is then compared to a background probability in order to cut any galaxies that are not likely to belong to the group. We note that there is no distinction made between groups and clusters in these catalogues. 
	
	Final halo masses are then calculated as follows: For groups with more than one member, the 
	\enquote{GAP correction} of \citet{2015MNRAS.450.1604L} is applied. This uses the luminosity difference between the brightest galaxy assigned to the group and then $n^{\mathrm{th}}$ nearest neighbour. For galaxies without companions, halo masses are estimated either using galaxy luminosity or stellar mass. We use the luminosity derived halo masses, which are derived from the relation between luminosity and halo mass yielded by the EAGLE simulations, defined as the total mass contained within a sphere where the density is 180 times the mean density of the Universe. \citet{2017MNRAS.470.2982L} state that the typical uncertainty of these estimations is $\sim 0.2$ dex. For the purposes of this analysis, we scaled the halo masses using the same cosmological parameters as in TNG300 for consistency between the observational data and simulations, whereby $h = 0.6774$ and $\Omega_\mathrm{m}$ = 0.3089 \citep{2016A&A...594A..13P}. It should be noted, however, that for all that concerns the shape and properties of the stellar-to-halo mass relations of the SDSS and TNG300 data, differences may arise because of modeling differences between the EAGLE and the TNG300 simulations and do not necessarily reflect failures of one or the other with respect to reality, which we cannot assess.
	
	In order to obtain further galaxy properties that were not contained in the \citet{2017MNRAS.470.2982L} catalogues, we cross-matched the sample with the New York University value-added catalogue of SDSS galaxies presented in \citet{2005AJ....129.2562B} and the \citet{2007ApJ...671..153Y} group catalogue, using the SDSS identification number. We then obtained properties such as the Malmquist bias weightings, which account for the sensitivity of the survey, allowing us to correct any average galaxy properties for galaxies that fall under the survey limits \citep[see][and references therein]{2007ApJ...671..153Y} and the S\'{e}rsic index (obtained by fitting azimuthally averaged surface brightness profiles), which provides an indicator for the characteristic galaxy morphology. All observational values presented here forth are corrected for these bias weightings.
	
	To link the star forming histories back to present day properties such as halo mass, we then further cross-matched the catalogues with those of \citet[][presented in Section~\ref{data:SED}]{2016ApJ...824...45P} using a matching radius of 5 arcsec. This reduces the likelihood of miss-matches while still allowing for small differences in the astrometry. We retrieved better constrained estimations of the stellar mass from the photometric SED fitting method of \citet[][described in Section~\ref{data:SED}]{2016ApJ...824...45P} than those in the other group catalogues, hence these are the stellar masses referred to here forth. We also used the specific star formation rates (sSFR) estimated by the photometric SED fitting and the {\it g} and {\it r} magnitudes used by \citet{2016ApJ...824...45P} for consistency.
	
	We finally applied a cut in halo mass of M$_{\mathrm{Halo}}$ $> 10^{12}$ M$_{\odot}$ and selected only central galaxies in their respective groups. This halo mass cut was chosen as below this value the halo mass estimates have significant uncertainties. Our final sample, hereafter referred to as our SDSS sample, contains 89,647 galaxies, of which 12,663 have companions according to the group catalogues and are hereafter referred to as group environments and 76,984 are without a companion, hereafter referred to as field galaxies. The sample spans a redshift range of $z = 0 - 0.13$. We characterise this sample in Figure~\ref{fig:properties}, where galaxy properties are plotted in 0.1 dex bins of stellar and halo mass. By splitting our sample into bins with a size of 0.1 dex in both stellar mass and halo mass and calculating the (Malmquist weighted) properties for each bin, we can reveal any trends in the data. This binning allows us to simultaneously control for both secular processes, of which stellar mass is a proxy and environmental processes, of which halo mass is a proxy. This enables us to highlight how secular and environmental processes can shape the evolution of a specific galaxy population, imprinted in their characteristic present day properties and average evolutionary sequences.
	
	The top left hand panel of Figure~\ref{fig:properties} shows the logarithm of the number of galaxies per bin of stellar and halo mass. As could be expected, those bins in the centre of the distribution contain the highest number of galaxies ($\gtrsim$ 1000 galaxies per bin) with the edges of the distributions containing the least ($\sim 10$ galaxies per bin). The top right hand panel shows the mean amount of satellites within the parent halo \citep[as determined by the modified FoF halo finder algorithm from][]{2017MNRAS.470.2982L}. The most massive galaxies (M$_{\rm Stellar}\gtrsim10^{11}$M$_{\odot}$) in the most massive haloes (M$_{\rm Halo}\gtrsim10^{13.5}$M$_{\odot}$) in our sample have the most satellites.
	
	The mean aperture size used in the photometry from \citet{2016ApJ...824...45P} divided by the SDSS effective radius from \citet{2005AJ....129.2562B} is plotted in the mid left panel. We see that most galaxies use an aperture of 1.9 effective radii with no obvious trend in halo or stellar mass, except for a sub-sample (M$_{\rm Stellar}\sim10^{11}$M$_{\odot}$, M$_{\rm Halo}\sim10^{12}$M$_{\odot}$) which have $\sim 0.3 R/R_e$ smaller aperture sizes. In the mid right panel we compute the mean S\'{e}rsic index (mid left panel) in order to give an indication of the average morphology of each bin. Similar to previous studies, we find that galaxies with greater stellar masses have, on average, higher S\'{e}rsic indices, with the exception of a sub-sample (M$_{\rm Stellar}\sim10^{11}$M$_{\odot}$, M$_{\rm Halo}\sim10^{12}$M$_{\odot}$) which show lower S\'{e}rsic indices at higher stellar masses for constant halo mass. We also see subtle halo mass dependencies, whereby in galaxies with stellar mass M$_{\rm Stellar} < 10^{11}$M$_{\odot}$, at constant stellar mass but with higher halo masses have, in general lower lower S\'{e}rsic indices. 
	
	The bottom left panel shows the mean {\it g} - {\it r} colour from the photometry used by \citet{2016ApJ...824...45P}. We see that the average colour of a galaxy correlates heavily with the stellar mass, however similar to the S\'{e}rsic index there is a slight halo mass dependence at stellar masses M$_{\rm Stellar} < 10^{11}$M$_{\odot}$. A galaxy of the same stellar mass in a more massive halo is, on average, bluer in this regime. The bottom right panel shows the mean sSFR as determined from the SED fitting of \citet{2016ApJ...824...45P}. The behaviour correlates in the same way as the S\'{e}rsic index, as a strong function of stellar mass, however with a secondary halo mass dependency below M$_{\rm Stellar} < 10^{11}$M$_{\odot}$, whereby galaxies with the same stellar mass but in more massive haloes have higher sSFRs. We also see an exception in a sub-sample of galaxies with M$_{\rm Stellar}\sim10^{11}$M$_{\odot}$ and M$_{\rm Halo}\sim10^{12}$M$_{\odot}$, which have higher sSFRs than less massive galaxies at constant halo mass.

	\subsection{The photometric SED fitting process}
	\label{data:SED}
	
	In their paper, \citet{2016ApJ...824...45P} carried out photometric SED fitting on a large sample of galaxies ($\sim$ 230,000) from SDSS. This catalogue of galaxies includes estimates of the assembly times (i.e. the formation time of both in- and ex-situ stars) of 10, 50 and 90 per cent of the stellar mass ($t_{10}$, $t_{50}$, $t_{90}$) as well as estimates of the Star Formation Rate (SFR) and stellar mass of each galaxy. A full description of the SED fitting process can be found in \citet{2012MNRAS.421.2002P} or \citet{2016ApJ...824...45P}, however we provide a brief outline here.
	
	\citet{2016ApJ...824...45P} initially take photometry in 4 optical bands from SDSS DR10 \citep{2014ApJS..211...17A}: {\it g}, {\it r}, {\it i} and {\it z} (central wavelength $\lambda_c$: 4770 \AA, 6231 \AA, 7625 \AA\ and 9134 \AA), excluding the {\it u}-band due to systematic differences of 0.1 mag compared to the model library \citep[see][for more details]{2016ApJ...824...45P}. The SDSS Petrosian radius is used to measure the photometric fluxes and has a median aperture of 5 arcsec, with 16$^{\mathrm{th}}$ and 84$^{\mathrm{th}}$ percentiles of 3.8 and 7.8 arcsec. In order to more accurately constrain the stellar masses, SFRs and the Star Formation Histories (SFHs), photometry was added in the ultraviolet from {\it GALEX}: the far-ultraviolet ($\lambda_c$ $\sim$ 1550 \AA) and the near-ultraviolet ($\lambda_c$ $\sim$ 2200 \AA), and in the infra-red from {\it WISE}: {\it W1} ($\lambda_c$ = 3.4$\mu m$). A 3 arcsec matching radius was applied between surveys. 
	
	All data were corrected for foreground extinction using the dust maps of \citet{2011ApJ...737..103S} and the extinction law of \citet{1999PASP..111...63F}, except for {\it W1}, where the reddening is assumed to be negligible. The redshifts required for the SED fitting process were obtained from optical fibre spectroscopy from SDSS DR10. \citet{2016ApJ...824...45P} also exclude any type 1 Active Galactic Nucleii (hereafter AGN), which are expected to significantly affect the shape of the SED, by using the emission line catalogue of \citet{2011ApJS..195...13O}. \citet{2016ApJ...824...45P} state they do not exlude type 2 AGN as previous studies argue that continuum emission from these objects does not affect the estimates of the physical parameters derived from optical photometric fits \citep{2003MNRAS.346.1055K}.
	
	Individual galaxies can undergo an extremely wide range of evolutionary histories, which in turn affects their SFH. This SFH is reflected in the shape of the expected SED, implying that a wide variety of SFHs needs to be constructed to compare to the observed photometry. The library of SFHs used in \citet{2016ApJ...824...45P} is built from the Millenium Simulation \citep{2005Natur.435..629S} post-processed with the semi-analytic models from \citet{2007MNRAS.375....2D}. These SFHs contain a number of different star formation scenarios (e.g bursty, smooth, declining etc.) and account for a number of different metal enrichment histories. 
	
	These models are then combined with stellar population synthesis models as described in \citet{2003MNRAS.344.1000B}. Nebular emission is computed using {\sc cloudy} \citep{1998PASP..110..761F, 2017RMxAA..53..385F} and dust attenuation models account for internal reddening, including both spatial distribution and orientation uncertainties, via an implementation of \citet{2000ApJ...539..718C}. The full library contains around 1.5 million possible SFHs and SED models.
	
	A Bayesian method is then applied to match the observed photometry of any one galaxy to the SED models. The weighted likelihood of all possible models is used to construct a probability distribution (median, 16$^{\mathrm{th}}$ percentile and 84$^{\mathrm{th}}$ percentile values) for host galaxy properties such as the stellar mass and SFR averaged over the last 10 Myr. The SFH, however, is calculated from the weighted 10 best model fits to significantly reduce computational time, this then yields the $t_{10}$, $t_{50}$ and $t_{90}$ values. We note that the SED fitting does not depend on the steepness of the stellar mass - halo mass relation and does not depend on the SAM models used \citep{2012MNRAS.421.2002P}. 
	
	The mean uncertainties on the assembly times range from 0.3 Gyr for $t_{90}$ to up to 0.9 Gyr for $t_{10}$, with mean uncertainties on the stellar mass and SFR estimations of 0.1 and 0.5 dex respectively. \citet{2016ApJ...824...45P} show that although a wider range or the full library of models could be used to generate better constrained probability distributions for the SFHs, the difference in the precision compared to using the 10 best fitting models does not justify the significant increase in computational time needed.
	
	As our SDSS sample is distributed between $0 < z < 0.13$, each lookback time is not the expected $z = 0$ value. We calculate the luminosity distance based on the redshift and apply a correction to the lookback time, to bring these values into line with those expected as if each galaxy was observed at $z = 0$. This provides a uniformity over the observational sample and allows an easier comparison with the simulations.
	
	\section{Simulation data and methods}
	\label{sec:Simulations}
	
	\subsection{IllustrisTNG}
	\label{sim:IllustrisTNG}
	
	Cosmological simulations have recently progressed to the stage where they can re-produce multiple properties of the observed universe to a significant degree of detail and accuracy. Some of the numerous examples include the reproduction of the stellar mass function, Hubble sequence of morphologies, colour bi-modality, star forming main sequence, AGN galaxy properties etc. \citep{2015MNRAS.446..521S, 2015MNRAS.450.4486F, 2017MNRAS.468.3395M, 2017MNRAS.470..771T, 2018MNRAS.475.1288S, 2018MNRAS.473.4077P, 2018MNRAS.475..676S, 2018MNRAS.474.3976G}. As our observational sample contains $\sim 90,000$ galaxies, IllustrisTNG provides the advantage of having a comparable sample size after matching in their 300 cMpc simulational run compared to other similar simulations which have sizes of 100 cMpc such as EAGLE or Horizon-AGN \citep{2014MNRAS.444.1453D}.
	
	IllustrisTNG is a suite of magnetohydrodynamical simulations \citep{2018MNRAS.475..648P, 2018MNRAS.480.5113M, 2018MNRAS.475..676S, 2018MNRAS.475..624N, 2018MNRAS.477.1206N} which build upon the original Illustris simulations \citep{2014MNRAS.444.1518V, 2014Natur.509..177V, 2014MNRAS.445..175G} while trying to address and improve on points of tension with observations \citep{2015A&C....13...12N}. It includes a number of different simulation boxes (50 co-moving Mpc, 100 cMpc and 300 cMpc per side), and dark-matter-only counterparts. The simulation uses the AREPO moving mesh code \citep{2010MNRAS.401..791S}, which utilises an adaptive Voronoi tessellation to solve the ideal magnetohydrodynamical equations and TreePM to solve the coupled equations of self gravitation at discrete timesteps. Galaxy groups are identified using a FoF algorithm, with galaxies identified using the \textsc{Subfind} algorithm \citep{2001MNRAS.328..726S} and connected in each timestep using merger trees.
	
	IllustrisTNG includes schemes to account for a number of astrophysical processes including some that take place below the resolution limit of the simulations, i.e. the sub-grid physics. This covers gas radiative processes featuring multiple heating and cooling mechanisms as well as star formation given by a pressure dependent law according to the models of \citet{2003MNRAS.339..312S}. Stellar evolution and chemical enrichment are governed by yield tables and stellar, supernova and AGN feedback are included \citep[see][and references therein]{2018MNRAS.473.4077P}. One of the main refinements in IllustrisTNG compared to the original Illustris simulations is the implementation of a different AGN feedback scheme. This includes both a thermally injected mode (at high Eddington accretion ratios) and a kinetically injected mode \citep[at low Eddington ratios, see][]{2017MNRAS.465.3291W}. 
	
	The sub-grid physics choices are made so that specific relations are approximately consistent with observations. These include the galaxy stellar mass function, the galaxy stellar mass - size relation, the stellar mass - black hole mass relation and the halo gas fraction versus halo mass at $z = 0$ as well as the cosmic star formation rate density evolution.
	
	\begin{figure}
		\centering
		\includegraphics[width=1.1\columnwidth]{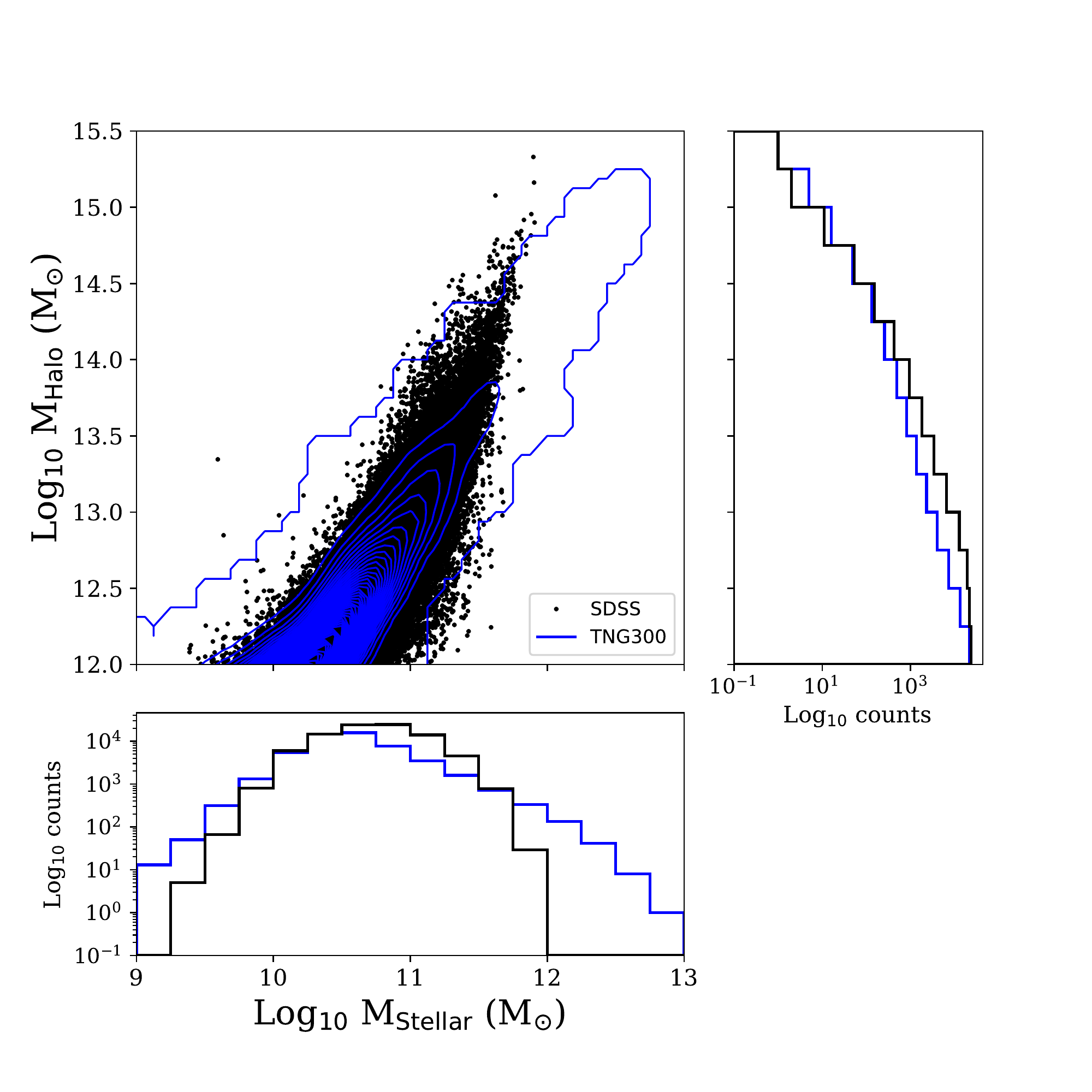}
		\caption{Central panel: The distributions in stellar mass versus halo mass for our SDSS observational sample (black points) compared to the simulated sample from TNG300 (blue contours). The 2 sub-panels show the respective distributions in log space of the two samples in stellar and halo mass in their corresponding colours.}
		\label{fig:sample}
	\end{figure}
	
	These improvements in the physics and sub-grid physics yield a better reproduction of a number of observables than the original Illustris simulations. \citet{2018MNRAS.475..648P} showed that IllustrisTNG reproduces galaxy sizes and the stellar mass function more accurately, especially at lower stellar masses. They also showed that the stellar-to-halo mass relation is more consistent with other semi-empirical findings and that the gas fraction to halo mass relation is in better agreement with X-ray inferences above a halo mass of $\sim 10^{13}$ M$_\odot$. \citet{2018MNRAS.475..624N} showed that IllustrisTNG reproduces the colour bi-modality to a better degree than the original Illustris, and \citet{2018MNRAS.475..676S} showed  that the clustering of red versus blue galaxies is also consistent with observations at $z \sim 0$.

	\subsection{Selection of the simulated sample}
	\label{sim:matching}
	
	To be able to make a meaningful comparison between observations and simulations, some level of matching needs to be applied. Firstly, the samples of galaxies in the considered data sets need to be selected as similarly as possible. Secondly, the galaxy properties from the simulated data need to be computed in as similar a way as possible to how they are inferred from the observational data. If the two data sets are unmatched then biases in either data set cause a comparison to be of little worth and the possibility that results appear completely different. A number of possible matching or selection techniques are in volume, stellar mass, redshift or flux. Some of these options, along with their possible impact on our results, are discussed further in Section~\ref{disc:limits}.
	
	In this paper, we use the publicly-available data from the largest simulation box \citep{2019ComAC...6....2N} with the best resolution at this level, TNG300-1 (hereafter TNG300), in order to obtain a simulated sample with a comparable number of galaxies to the SDSS set and thereby reduce volume-driven biases. We use the $z=0$ snapshot from the TNG300\footnote{we correct SDSS stellar assembly times to bring them in line with expectations at $z=0$, see Section~\ref{data:SED}} and apply a cut in halo mass, computed as the total mass of all particles enclosed in a sphere whose mean density is 200 times the mean density of the universe, to all galaxies of M$_{\mathrm{Halo}}$ > 10$^{12}$ M$_{\odot}$. We note that although this radius is not 180 times the mean density which is used for the observations, it is the closest match in the TNG catalogues, and that using subtly different halo mass definitions has a minimal effect on later results.
	
	We then select all central galaxies above the halo mass cut defined above. This yields 46,241 central galaxies, hereafter our TNG300 galaxy sample. We also note that although the FoF halo finders used for the simulations and the observational group catalogues are not exactly the same, such as the designation of the central galaxy in a halo being the most luminous member of the assigned group in the observations compared to the most massive in the simulations, we proceed by assuming that the algorithms should provide similar results as they are based on the same basic concepts, and any minor errors should not significantly impact trends in the data as we expect this to be mitigated by binning in both stellar and halo mass. A thorough test of this, however, is beyond the scope of this study.
	
	The majority of the SDSS photometry used in \citet{2016ApJ...824...45P} was measured within approximately two effective radii (2$R_e$, see Figure~\ref{fig:sample}), hence we choose to use this as our matching criterion for the simulations. Although the photometry apertures are not uniform and there is a slight scatter we do not expect this to impact significantly on the results and trends we see. We take the available aperture of 2$R_{half-mass}$ in TNG300 (in three dimensions) to calculate the majority of our values. We stress at this point that we are making the assumption that $R_{half-mass}$ and $R_{e}$ are equivalent, which although not exact \citep{2019arXiv190410992S, 2019MNRAS.490.3196P} is not expected to significantly impact our results.  All measurements from the simulation are here forth within the three-dimensional stellar half-mass radius (2$R_{half-mass}$).
	
	As explained in Section~\ref{data:SED}, the $t_{10}$, $t_{50}$ and $t_{90}$ times for the observational data are calculated from the weighted 10 best-fitting SFHs. These SFHs are also weighted by the initial stellar birth mass. In order to create the best comparison possible, we considered all the present day stellar particles in each TNG300 galaxy within 2 $R_{half-mass}$. We then used the formation times of each stellar particle weighted by the initial stellar birth mass of each particle (the same as the observations) to construct a profile of the cumulative initial stellar mass at birth as a function of time for each galaxy in the simulation. From these profiles we calculated the corresponding $t_{10}$, $t_{50}$ and $t_{90}$ times for each galaxy in our simulated sample.  
	
	We also retrieve the stellar masses which are calculated from the combined mass of all stellar particles within 2$R_{half-mass}$, and are re-scaled according to \citet{2018MNRAS.473.4077P} to account for biases caused by the lower resolution limit of the simulation with a 300 cMpc box size (TNG300) compared to the simulation with higher resolution with a 100 cMpc box (TNG100). We compute the SFR averaged over the last 10 Myr \citep{2019MNRAS.485.4817D} within 2$R_{half-mass}$. This provides a comparable estimation to the observations, which are also averaged over 10 Myr. In their study \citet{2019MNRAS.485.4817D} show that resolution has no significant effect on the SFRs, therefore no re-scaling of the SFRs due to resolution is needed. Colours, namely the {\it g} and {\it r} magnitudes, are calculated from summing the luminosity of all stellar particles belonging to that galaxy within 2$R_{half-mass}$. We also convolve all simulational values with the uncertainties in the observational results to better represent the simulational data. 
	
	A comparison of the final distributions in stellar versus halo mass of the observational sample and simulated sample can be seen in Figure~\ref{fig:sample}. We highlight again that SDSS galaxies halo mass estimates are generated from mock catalogues of the EAGLE simulation and therefore no conclusions can be derived here regarding the level of realism of the TNG300 model in relation to the overall shape of its stellar-to-halo mass relation. Figure~\ref{fig:sample} shows that the distributions in halo masses are similar, however our sample from IllustrisTNG has a slightly different stellar mass distribution, compared to our observational sample. These differences however are small at the most massive end ($<$ 100 galaxies with M$_{\rm Stellar}\gtrsim10^{12}$M$_{\odot}$) and should be somewhat accounted for by binning in both stellar and halo mass, minimising the impact on the average trends.

	\section{Results}
	\label{sec:Results}
	
	\subsection{Average assembly times of SDSS and TNG300 galaxies}
	\label{res:formation times}
	
\begin{figure*}
		\centering
		\includegraphics[width=1.5\columnwidth]{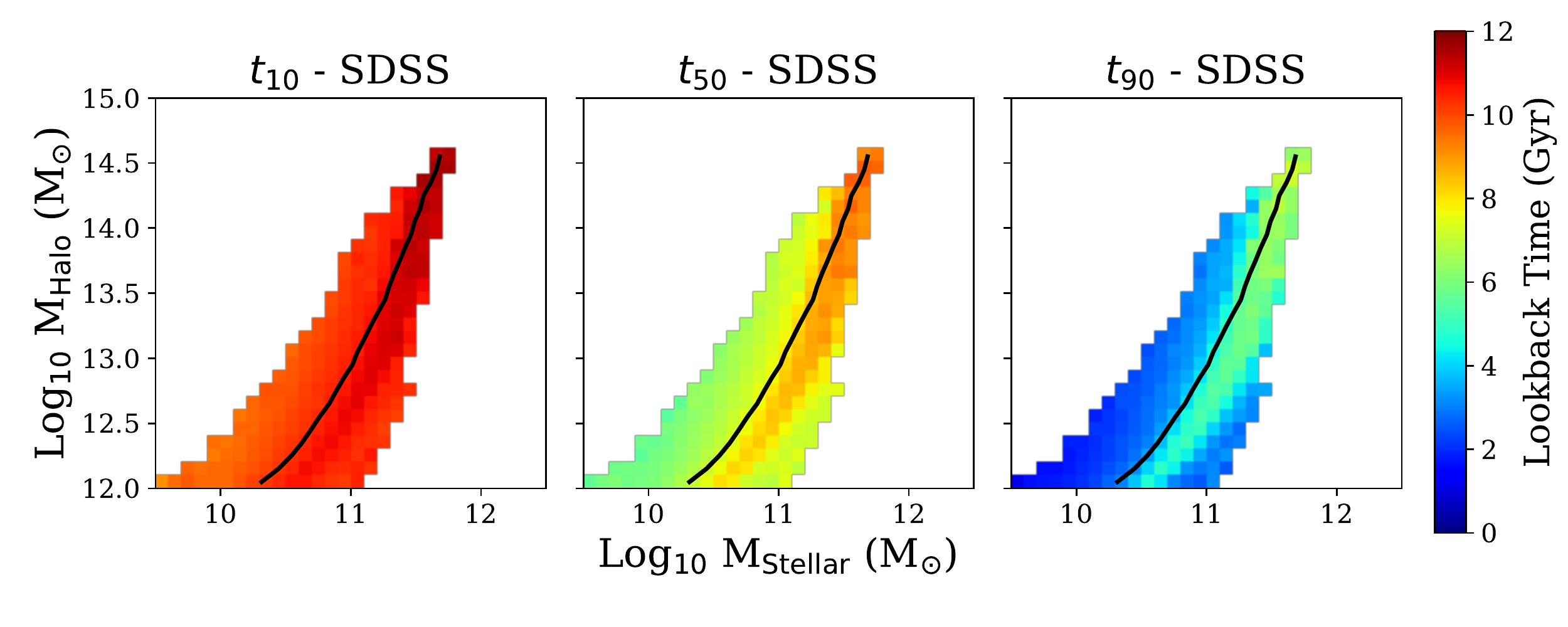}
		\caption{The mean lookback times at which 10\% (left hand panel), 50\% (central panel) and 90\% (right hand panel) of the stellar mass was assembled for our SDSS sample of central galaxies at $z=0$ in bins of halo and stellar mass. The colour bar gives the lookback time at which these percentages were formed. Going from the left to right of each distribution, we initially see clear downsizing behaviour, whereby more massive galaxies assembled specific percentages of their stellar mass at earlier times. An exception to this trend is a sub-population in the bottom right of the distribution which can be seen to display the opposite trend, i.e. more massive galaxies at fixed halo mass with younger ages. The solid black lines give the median stellar-to-halo mass relation of the sample.}
		\label{fig:av_times}
	\end{figure*}

	To compare at which epochs our observational and simulated samples assembled most of the stellar mass in their inner regions and how environmental and secular processes may shape these processes, we initially investigate the stellar mass assembly times. As outlined in Section~\ref{data:SED}, the SED fitting process yields estimations for the lookback time at which each galaxy in our observational sample formed 10, 50 and 90 per cent of its stellar mass in the SDSS sample. Like in Figure~\ref{fig:properties}, we bin in 0.1 dex bins of stellar and halo mass. The results are presented in Figure~\ref{fig:av_times}. 
	
\begin{figure*}
		\centering
		\includegraphics[width=1.5\columnwidth]{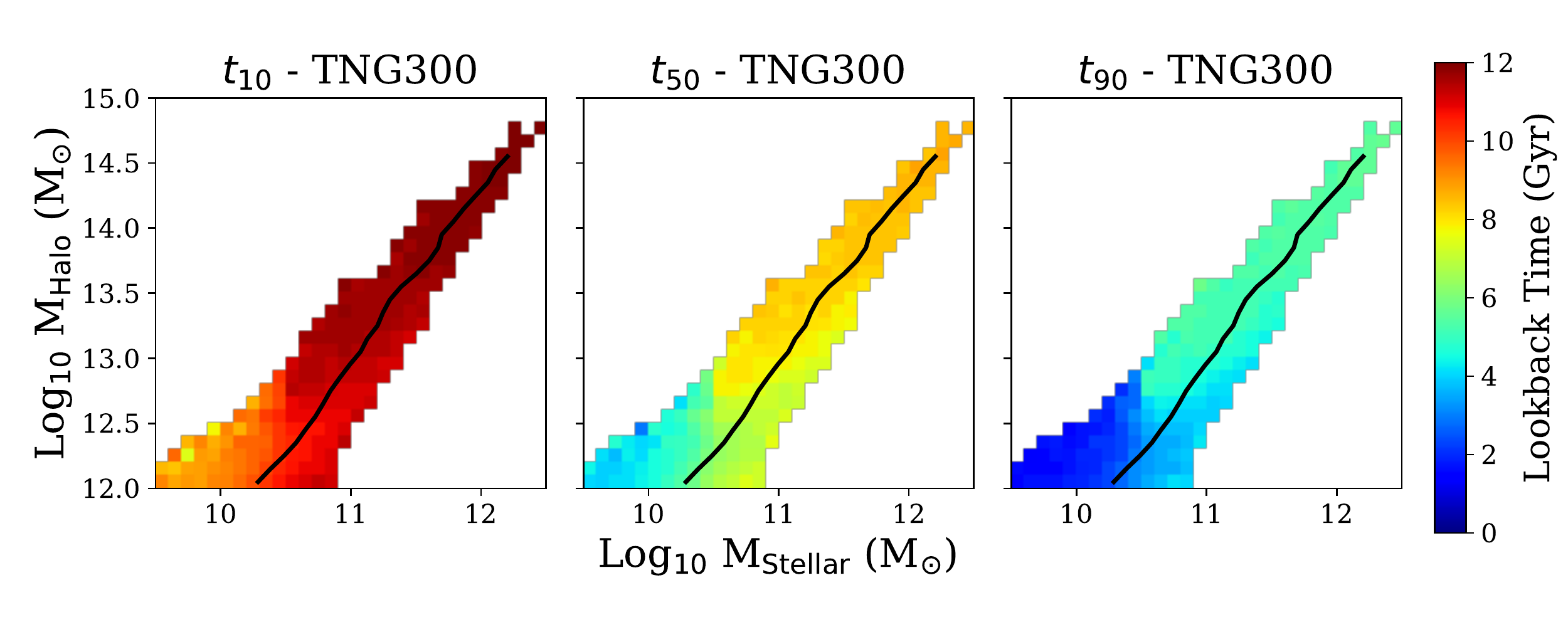}
		\caption{The mean assembly times at which 10\% (left hand panel), 50\% (central panel) and 90\% (right hand panel) of the stellar mass was assembled for our TNG300 sample of central galaxies at $z=0$ in bins of present day halo and stellar mass. The colour bar gives the lookback time at which these percentages were formed. The simulation predicts lookback times, in general, that are similar to the observations, however it predicts a different overall behaviour with respect to halo and stellar mass than the observations. The solid black lines give the median stellar-to-halo mass relation of the sample.}
		\label{fig:ill_av_times}
	\end{figure*}

	We also test how well TNG300 reproduces the assembly of 10, 50 and 90 per cent of the stellar mass as a function of present day stellar and halo mass by applying the same binning procedures as for the observational data, to provide a like for like comparison: results are given in Figure~\ref{fig:ill_av_times}.
	
	Figure~\ref{fig:av_times} shows the mean lookback times at which 10 (left hand panel), 50 (central panel) and 90 (right hand panel) per cent of the stellar mass was assembled. We see that the range of assembly times in $t_{10}$ is relatively small, ranging from 12 Gyr to 10 Gyr. This range increases as we move to $t_{50}$ (6 - 9 Gyr) and to $t_{90}$ (2-6 Gyr). There is also a strong dependency of the lookback times on stellar mass, whereby holding the halo mass constant (i.e going from left to right) we observe, on average, a clear case of downsizing. This is in line with previous studies of galaxy assembly \citep[see e.g.][]{2005ApJ...621..673T, 2005MNRAS.362...41G, 2010MNRAS.407..937P, 2016ApJ...824...45P}, whereby massive galaxies form a certain percentage of their stars earlier than less massive galaxies. We also see, however, a simultaneous dependency on the halo mass, whereby galaxies at constant stellar mass (below a stellar mass of 10$^{11}$ M$_\odot$) but with higher halo masses, display on average younger lookback times, especially $t_{50}$ and $t_{90}$. This reflects the behaviour seen in the sSFRs and S\'{e}rsic indices from Figure~\ref{fig:properties}.
	
	A sub-sample ($\sim$2 per cent of our SDSS sample), located in the bottom right corner of the distribution, seems to contradict hierarchical mass assembly behaviour. These galaxies have the highest stellar-to-halo mass ratios (M$_{\mathrm{Stellar}} \sim$ 10$^{11}$ M$_\odot$ and M$_{\mathrm{Halo}} \sim$ 10$^{12}$), and have younger ages than some less massive galaxies at constant halo mass. This is the same sub-sample which have low S\'{e}rsic indices and raised sSFRs in Figure~\ref{fig:properties}. We investigate this sub-population further in later sections and discuss these results in Section~\ref{disc:observations}.
	
	We see that the average assembly times are, in general, fairly well reproduced by TNG300, as seen in Figure~\ref{fig:ill_av_times}. Our TNG300 sample generally predicts similar times for the assembly of 10 (left hand panel), 50 (central panel) and 90 (right hand panel) per cent of the stellar mass across most stellar and halo masses. There are some subtle differences, such as at the lowest stellar masses (M$_{\mathrm{Stellar}} \sim$ 10$^{10}$ M$_\odot$) in the $t_{10}$ and $t_{50}$ times, which tend to be different by $\sim$ 0.5 - 1 Gyr, however these are still within the observational uncertainties. We also see that although the most massive objects in general formed early on in TNG300, the same as in the observations, a significant fraction of the oldest assembled objects appear to have intermediate stellar and halo masses (M$_{\mathrm{Stellar}} \sim$ 10$^{10.5}$ M$_\odot$ and M$_{\mathrm{Halo}} \sim$ 10$^{13}$). This is in qualitative disagreement with our observational results, where these objects usually have some of the youngest assembly times. These differences in the lookback times can be up to 2-3 Gyr, well outside of the observational errors. We discuss these differences in more detail in further sections.
	
	\subsection{Differences in the lookback times between SDSS and TNG300}
	\label{res:ill_limit}

	\begin{figure*}
		\centering
		\includegraphics[width=1.5\columnwidth]{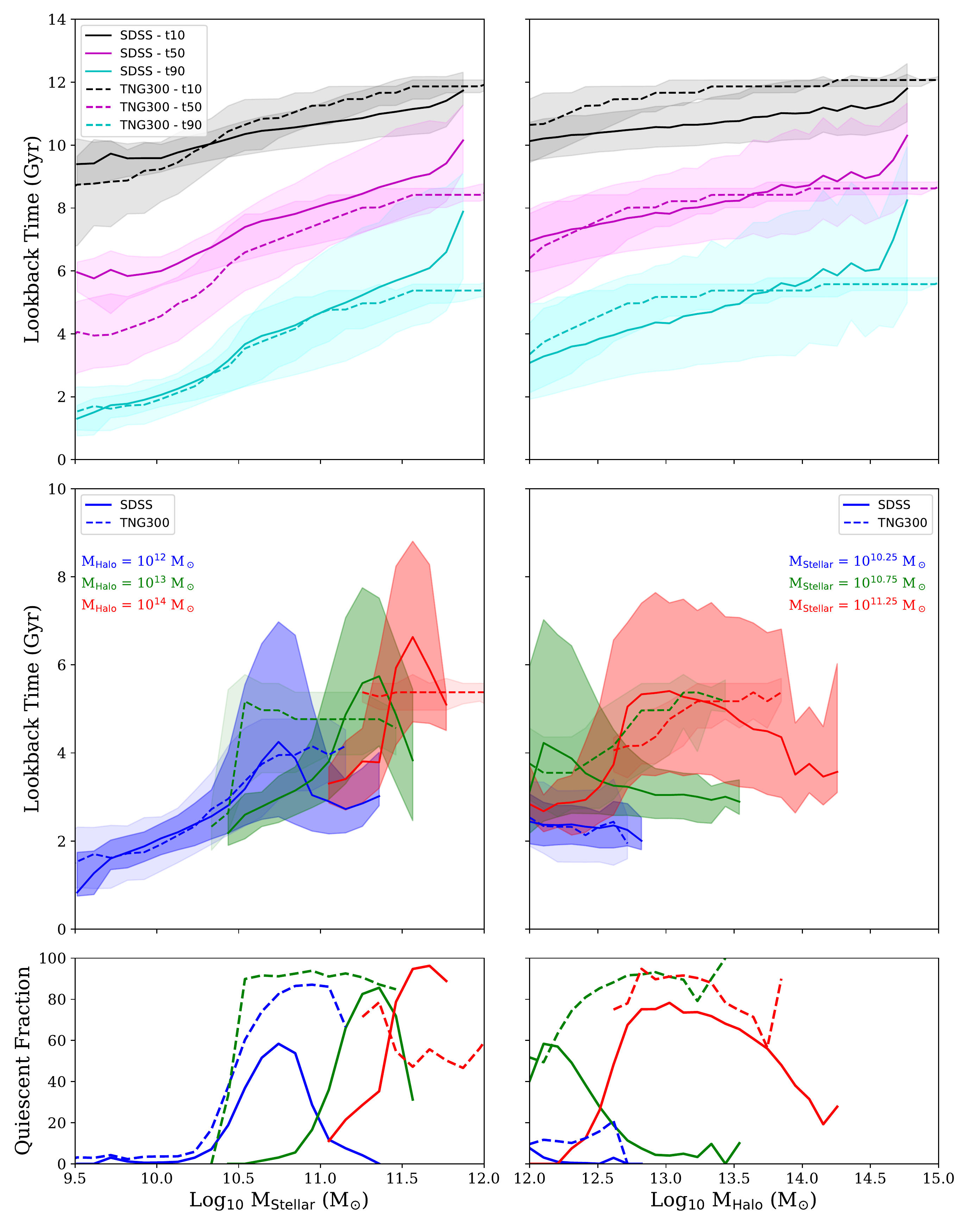}
		\caption{Top row: The $t_{10}$ (black), $t_{50}$ (magenta) and $t_{90}$ (cyan) times for our observational sample (solid line) compared to TNG300 (dashed) as a function of stellar mass (left panel) and halo mass (right panel). We see generally good agreement, whereby TNG300 is mostly within the scatter of the observational distribution. Middle row: The $t_{90}$ lookback times as a function of stellar mass (left panel) and halo mass (right panel), split into three different halo/stellar masses. When we account for both halo and stellar mass we see disparities between the observations and simulations, especially at intermediate values. Bottom row: The fraction of galaxies that are quiescent (Log$_{10}$ sSFR < -11 Gyr$^{-1}$) as a function of stellar and halo mass, split into the bins in halo/stellar mass, respectively, as above.}
		\label{fig:halo_comp}
	\end{figure*}
	
	Having qualitatively presented the stellar assembly times of both our SDSS and TNG300 samples in Section~\ref{res:formation times}, we now attempt to quantify and compare some of these results. In order to do this we firstly compare trends in either halo or stellar mass to see how well TNG300 reproduces the SDSS data if only one variable is considered. This is presented in the top row of Figure~\ref{fig:halo_comp}. The trends of $t_{10}$, $t_{50}$ and $t_{90}$ as a function of stellar mass alone are in left hand panel and halo mass alone in the right hand panel. The solid lines give the medians of the SDSS data with 16$^{\mathrm{th}}$ and 84$^{\mathrm{th}}$ percentiles given by the shaded area and the dotted lines give the median of the TNG300 sample. We see that the trends in the observational sample can be fairly well reproduced by the simulation, as the simulated values are mostly within the scatter of the observational sample at almost all stellar and halo masses. We also, once again, see downsizing, whereby less massive galaxies have younger $t_{10}$, $t_{50}$ and $t_{90}$ values as a function of both stellar and halo mass. Comparing the top left panel to the top right panel we see that these trends tend to be stronger (represented by the steeper gradients) in stellar mass than in halo mass for all lookback times. We also note an increase in the scatter of the distributions from an average of $\sim 2$ Gyr in $t_{10}$ to $\sim 4$ Gyr in $t_{90}$, as seen above in section~\ref{res:formation times}.
	
	To quantify the effect that accounting simultaneously for stellar and halo mass has on any comparison, and thereby highlighting tensions between the simulations and observations, we took 3 different samples at low (M$_{\mathrm{Halo}}$ = 10$^{12}$ M$_\odot$), intermediate (M$_{\mathrm{Halo}}$ = 10$^{13}$ M$_\odot$) and high halo masses (M$_{\mathrm{Halo}}$ = 10$^{14}$ M$_\odot$) with a bin width of $\pm$ 0.2 dex around the central halo mass and plotted the median, 16$^{\mathrm{th}}$ and 84$^{\mathrm{th}}$ percentiles of the $t_{90}$ values as a function of stellar mass, seen in the left mid panel of Figure~\ref{fig:halo_comp}. We also take three stellar masses (M$_{\mathrm{Stellar}}$ = 10$^{10.25}$, 10$^{10.75}$, 10$^{11.25}$ M$_\odot$ $\pm$ 0.1 dex) and plotted the median and 16$^{\mathrm{th}}$ and 84$^{\mathrm{th}}$ percentiles of the $t_{90}$ values as a function of halo mass in the right mid panel. We use $t_{90}$ as this is the best constrained value of the three lookback times, with the smallest uncertainties. 
	
	We firstly observe a consistent, non-monotonic behaviour in the observational sample (median given by the solid line and 16$^{\mathrm{th}}$ and 84$^{\mathrm{th}}$ percentiles shaded) when considering the halo mass sub-samples as a function of stellar mass (mid left hand panel). We also see that for most stellar and halo masses, TNG300 (median given by the dashed line and 16$^{\mathrm{th}}$ and 84$^{\mathrm{th}}$ percentiles shaded) predicts lookback times within the scatter of SDSS sample. We see, however, that the greatest differences between the simulation and observations are at intermediate stellar and halo mass. This difference can be up to 3 Gyr, significantly outside of the observational errors. At these stellar masses and greater, we also observe subtly different trends compared to those inferred from observations, although the values are within the scatter of the distributions.
	
	In the mid right hand panel we see the lookback times as a function of halo mass in the 3 different stellar mass bins defined above. We see young lookback times with a flat trend as a function of halo mass for the lowest stellar mass bin. For intermediate stellar masses we see intermediate lookback times, with a slight negative trend between the lookback time and halo mass, i.e. galaxies in more massive haloes have slightly younger ages. This behaviour is repeated in the highest stellar mass bin, except at low halo masses, where the sub-population with the highest stellar-to-halo mass ratios lowers the lookback times. When we compare this to TNG300, we see a good agreement of the lookback times in the lowest stellar mass bin and a reasonable agreement in the highest stellar mass bin between the observations and simulations as a function of halo mass. We see, however, that the halo mass trends in TNG are reversed with respect to the observations in the intermediate and high stellar mass ranges (M$_{\mathrm{Stellar}}$ $\sim$ 10$^{10.75}$ M$_\odot$, 10$^{11.25}$ M$_\odot$), whereby $t_{90}$ increases with halo mass at fixed stellar mass. This causes differences of up to $\sim 3$ Gyr at fixed stellar and halo mass. We also observe that the scatter in M$_{\mathrm{Halo}}$ at fixed M$_{\mathrm{Stellar}}$ is smaller than in the observations.
	
	To further investigate the drivers behind these differences in the stellar assembly times when we account for both stellar and halo mass, we investigate the quiescent fraction of galaxies \citep[here defining quiesence as Log$_{10}$ sSFR < -11 yr$^{-1}$][]{2019MNRAS.484.1702P} as this can cause major differences when comparing the ages of two galaxy populations. The quiescent fraction of galaxies yielded by IllustrisTNG has been previously investigated, reproducing observational results to a good agreement at both $z=0$ and $z\lesssim2$ \citep{2019MNRAS.485.4817D}, however much like the stellar mass assembly times, we want highlight the effects that stellar and halo mass may have on sub-populations and reveal more subtle trends.
	
	In Figure~\ref{fig:halo_comp}, bottom panels, we see that TNG300 (dashed lines) reproduces the observed quiescent fraction (solid lines) well in the domains where the lookback times have some level of agreement in the mid panels, however in the domains where we have differences in the lookback times as both a function of stellar or halo mass, there are major differences in the quiescent fraction. This is best seen in the intermediate halo mass bin in the left hand panel at intermediate stellar masses, whereby TNG300 predicts that $\sim$80 per cent of all galaxies should be quiescent, while in the observations $<20$ per cent are, and the lookback times are therefore much higher than expected. This is confirmed by the intermediate stellar masses as a function of halo mass in the right hand panel, whereby nearly all galaxies in this sub-sample are quiescent compared to $<20$ per cent on average for the observations.
	
	\subsection{Characteristic galaxy properties in TNG300}
	\label{ill:properties}
	
	\begin{figure}
		\centering
		\includegraphics[width=1\columnwidth]{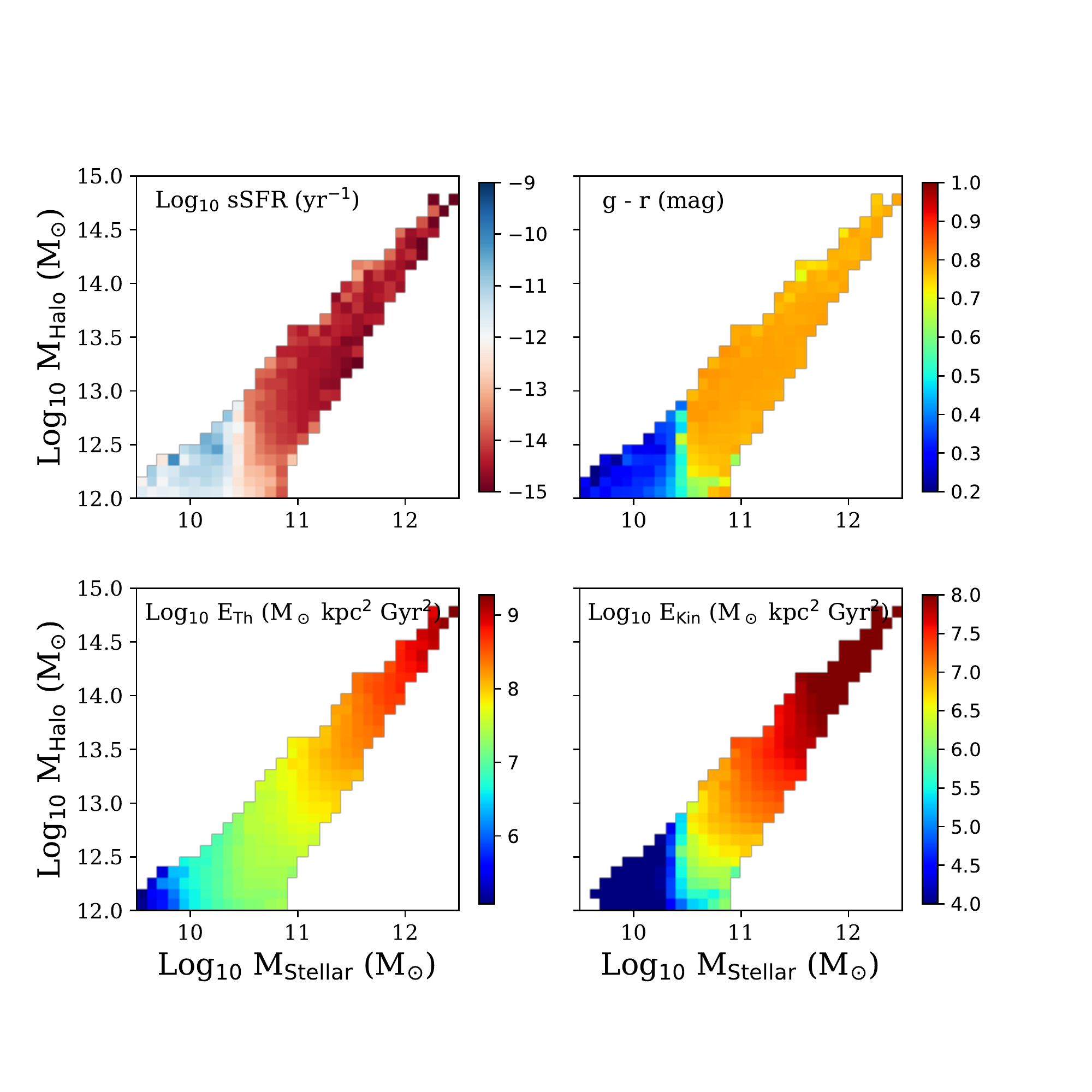}
		\caption{The top left panel shows the mean sSFR and the top right panel shows the mean $g - r$ colour as a function of both stellar and halo mass in TNG300. The bottom left panel shows the mean cumulative energy injected into the galaxy thermally from AGN feedback, and the bottom right panel shows the same but kinetically. The kinetic mode feedback reflects on average the trends seen in the upper panels and in the assembly times of galaxies in Figure~\ref{fig:ill_av_times}.}
		\label{fig:ill_sfr_col}
	\end{figure}
	
	To investigate possible mechanisms that may give rise to the tensions seen in the assembly times between SDSS and TNG300, especially at intermediate stellar masses (M$_{\rm{Stellar}}$ $\sim$ 10$^{10.5}$ M$_{\odot}$), we explored a number of host galaxy properties at $z = 0$ in TNG300 that were likely to be physically linked to the star formation histories or to the quenched fraction of galaxies. Any behaviour or trends in these properties as a function of both stellar and halo mass may indicate possible drivers of these differences in TNG300. Figure~\ref{fig:ill_sfr_col} shows the most relevant properties to this investigation. 
	
	In the top left hand panel we have the sSFR as a function of stellar and halo mass. We see a significant change in the sSFR at stellar masses of M$_{\rm{Stellar}}$ $\sim$ 10$^{10.5}$ M$_{\odot}$ with a very slight halo mass dependence at low halo masses (M$_{\rm{Halo}}$ $\sim$ 10$^{12 - 12.5}$ M$_{\odot}$). In the top right hand panel, we show the {\it g} - {\it r} colour binned in stellar and halo mass. The trends in the colours mirror the trends in the sSFR closely, whereby galaxies with lower sSFR rates have redder colours, as expected. The mean cumulative amount of energy injected into the galaxy from AGN feedback via the thermally injected mode (E$_{\mathrm{Th}}$) is shown in the bottom left hand panel, compared to the kinetically injected mode (E$_{\mathrm{Kin}}$) in the bottom right hand panel. We see that the amount of energy supplied by the thermal mode increases approximately with stellar mass relatively smoothly. The mean cumulative energy injected by the kinetic mode, however, shows more similar behaviour to the colours and sSFR at M$_{\rm Stellar}$ $\sim$ 10$^{10.5}$ M$_{\odot}$, with a sharper transition from low to high amounts of energy injected compared to the thermal mode, with a slight halo mass dependence after this point, reflecting the behaviour seen in the assembly times and sSFRs. 
	
	We postulate a scenario whereby galaxies in IllustrisTNG reach a critical stellar/halo mass (M$_{\mathrm{Stellar}} \sim$ 10$^{10.5}$ M$_\odot$, M$_{\mathrm{Halo}} \sim$ M$^{12.5}$ M$_\odot$) and the kinetic mode AGN feedback starts to take effect. Taking a scenario with two galaxies of similar halo mass but slightly different stellar masses, we propose that the AGN feedback may rapidly shut down star formation in the slightly less massive galaxy, resulting in old stellar assembly lookback times and higher quenched fraction as seen for intermediate mass galaxies (M$_{\mathrm{Stellar}} \sim$ 10$^{10.5}$ M$_\odot$, M$_{\mathrm{Halo}} \sim$ 10$^{13}$ M$_\odot$). This means the second galaxy with slightly more stellar mass is more resistant, due to a larger gravitational potential, and can continue forming stars, albeit likely at a lower rate, thereby lowering the stellar assembly lookback times. Beyond this critical stellar/halo mass ($\sim 0.5$ dex higher in stellar and halo mass), the kinetic mode AGN feedback dominates, shutting down star formation in most systems and causing lookback times to be even in stellar and halo mass phase space. Further investigation beyond the scope of this research is needed in order to confirm this scenario however.

	\subsection{Growth of the observational sample}
	\label{res:growth}
	
	\begin{figure*}
		\centering
		\includegraphics[width=1.6\columnwidth]{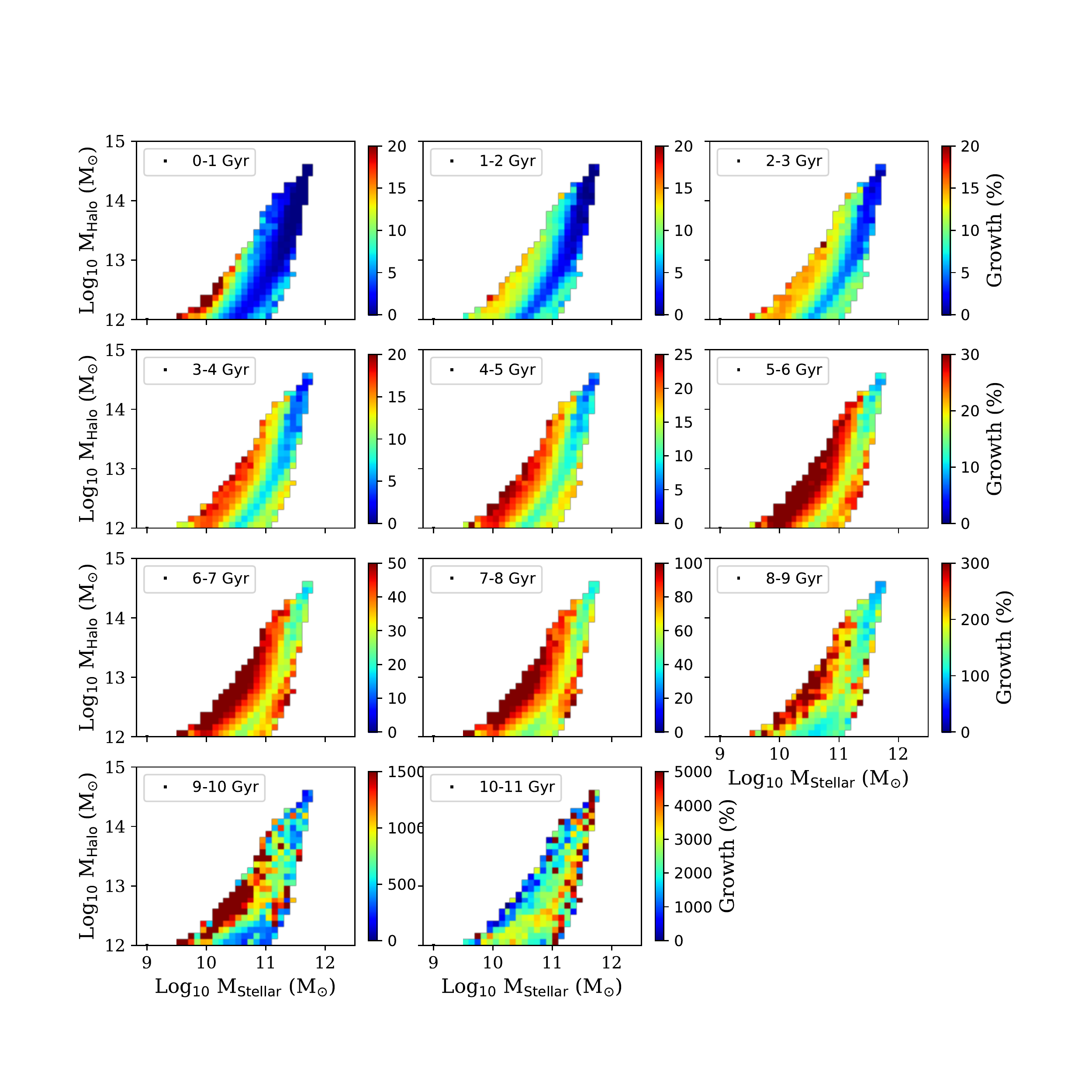}
		\caption{The average percentage growth rate of each SDSS galaxy bin over a Gyr (lookback time given in each panel) as defined in Equation~\ref{eq:growth}. We see that the lowest mass galaxies have significantly the highest rate of growth from $\sim$ 5 Gyr ago until the present day. We also see that the galaxies with high stellar-to-halo mass ratios have relatively steady growth rates at all epochs, although significantly less than the lowest mass galaxies from $\sim$ 5 Gyr ago to the present.}
		\label{fig:grow_pac}
	\end{figure*}
	
	In order to investigate the assembly of each present day stellar and halo mass bin of our observational sample in more detail, with a focus on the galaxy sub-population with the highest stellar-to-halo mass ratios (M$_{\mathrm{Stellar}} \sim$ 10$^{11}$ M$_\odot$ and M$_{\mathrm{Halo}} \sim$ 10$^{12}$) and with young $t_{90}$ lookback times, we calculate the mean growth rate of each galaxy per Gyr. This reveals at which epochs certain sub-populations of galaxies experienced greater rates of growth with respect to others, disclosing more detail about the evolution of each sub-population. As outlined in Section~\ref{data:SED}, the SED fitting process also estimates the stellar mass of each individual galaxy in 1 Gyr intervals, ranging from 0 Gyr to 11 Gyr. From this we can define the growth of a galaxy per Gyr as follows: 
	
	\begin{equation}
	Growth = \Delta M / M = (M_i - M_{i + 1}) / M_{i + 1}
	\label{eq:growth}
	\end{equation}
	
	where M$_i$ is the stellar mass of a galaxy at a specific epoch and M$_{i + 1}$ is the stellar mass of the same galaxy 1 Gyr earlier (e.g. when M$_i$ = M$_{\mathrm{Stellar}}$, 1 Gyr ago, M$_{i + 1}$ = M$_{\mathrm{Stellar}}$, 2 Gyr ago).
	
	Applying the same binning procedure in present day stellar and halo mass as before and calculating the Malmquist weighted means of each bin yields Figure~\ref{fig:grow_pac}. This shows the mean stellar mass growth of each bin in 1 Gyr intervals, from 11 Gyr ago up to the present day, with the specific epoch given by the labels inset in each plot. We note here that each sub-plot has a separate colour bar, in order to be able to differentiate growth accordingly at various epochs. We see that the average trends in stellar and halo mass in the observational sample from Figure~\ref{fig:properties} and Figure~\ref{fig:av_times} are replicated. At fixed M$_{\mathrm{Halo}}$ the mass growth	slows down as M$_{\mathrm{Stellar}}$ increases, while at fixed M$_{\mathrm{Stellar}}$ (where M$_{\mathrm{Stellar}} <$ 10$^{11}$ M$_\odot$) the mass growth is stronger in more massive haloes.
	
	The most massive galaxies in this sample (M$_{\mathrm{Stellar}} \sim$ 10$^{11.5}$ M$_\odot$, M$_{\mathrm{Halo}} \sim$ 10$^{14}$ M$_\odot$) have undergone, relatively, very little growth in stellar mass since early epochs, as expected from the old lookback times in Section~\ref{res:formation times}. The sub-population of galaxies with the highest stellar-to-halo mass ratios (M$_{\mathrm{Stellar}} \sim$ 10$^{11}$ M$_\odot$ and M$_{\mathrm{Halo}} \sim$ 10$^{12}$) and with young $t_{90}$ lookback times from Figure~\ref{fig:av_times}, appear to have had a fairly steady relative growth throughout cosmic time. The growth rate in stellar mass for this sub-sample in the last 6 Gyr is an average of $\sim$ 10 per cent per Gyr and attains a maximum of $\sim$ 25 per cent per Gyr. 
	
	We also see that those galaxies with low stellar mass and low stellar-to-halo mass ratios (i.e. those on the left hand side of the distribution) have the highest relative growth from $\sim$ 5 Gyr ago until the present day, as expected for low mass star forming galaxies. We note, however, that this growth rate is still at maximum $\sim$ 30 per cent per Gyr in the last 6 Gyr, and is unlikely to be high enough to shift them towards the median of the distribution in stellar and halo mass phase space. 
	
	\section{Discussion}
	\label{sec:Discussion}
	
	In this paper we have, for the first time, investigated the stellar mass assembly of $\sim 90,000$ low redshift, massive, central galaxies as a function of stellar and halo mass and linked their average assembly histories and characteristic present day properties to the environmental and secular processes that have influenced these characteristics. This was achieved by cross matching multiple group catalogues from SDSS \citep{2005AJ....129.2562B, 2007ApJ...671..153Y} with SED fitted star formation history estimations from \citet{2016ApJ...824...45P} and halo masses derived from mock catalogues from the EAGLE simulation \citep{2017MNRAS.470.2982L}. This observational sample was then compared to a sample from the TNG300 simulation in order to investigate if the simulations reproduce the observational trends. 
	
	\subsection{Trends in the formation times}
	\label{disc:observations}
	
	We have seen that a number of interlinked galaxy properties display the same average trends as a function of stellar and halo mass, such as the S\'{e}rsic index, sSFR, lookback times and growth rates. We will now discuss possible scenarios behind this behaviour. Figures~\ref{fig:properties}, \ref{fig:av_times}, and \ref{fig:grow_pac} show that all of these galaxy properties have a strong dependency on stellar mass. We see that, in general, more massive galaxies (at constant halo mass) assembled certain percentages of their stellar mass earlier. This result follows previous studies which find downsizing, whereby more massive objects assemble specific percentages of their stellar mass at earlier cosmic times than less massive objects \citep{1996AJ....112..839C, 2005ApJ...621..673T, 2016ApJ...824...45P}. 
	
	We also see, however, that there are secondary dependencies on halo mass. If we hold stellar mass constant (in the regime that M$_{\mathrm{Stellar}} <$ 10$^{11}$ M$_\odot$), those galaxies which reside in more massive haloes, tend to have younger lookback times which are reflected in their higher growth rates, higher sSFRs and lower S\'{e}rsic indices. One scenario could be that the more massive halo has a larger gravitational potential and is able to cool hot accreted gas more efficiently for steady star formation, presented as the "hot-mode" accretion in \citet{2003ASSL..281..185K} and \citet{2005MNRAS.363....2K} or recycle earlier accreted hot gas easier \citep[e.g.][]{2020arXiv200508995O}. A second scenario could be that higher mass haloes contain more satellites, which could supply gas to their centrals via interactions and mergers in order to sustain their star formation \citep[e.g.][]{2007MNRAS.375....2D, 2020MNRAS.494.5568J}.
	
	The exception to these observed behaviours is a sub-population of objects (M$_{\mathrm{Stellar}} \sim$ 10$^{11}$ M$_\odot$ and M$_{\mathrm{Halo}} \sim$ 10$^{12}$) with high stellar-to-halo mass ratios, which constitute $\sim$2 per cent of our entire sample and drive the non-monotonic trends seen in the lookback times when comparing different halo masses as a function of stellar mass (Figure~\ref{fig:halo_comp}). We postulate that these objects are massive spirals in the field with a relatively quiet accretion history. The characteristic low S\'{e}rsic indices indicate late-type morphologies, while the higher sSFRs show that star formation is not yet quenched in these systems, a phenomenon that is fairly rare at stellar masses of 10$^{11}$ M$_{\odot}$. According to the group catalogues, they are also located on average in haloes with no companions, supporting the idea that they may not have been morphologically transformed due to a lack of interactions, at least in recent times. To test this hypothesis, we also visually inspected 50 of these objects selected at random using SDSS imaging data, finding no significant signs of interactions, very few neighbours and mostly disk-like structure in the majority of these objects.

	\begin{figure*}
		\centering
		\includegraphics[width=1.8\columnwidth]{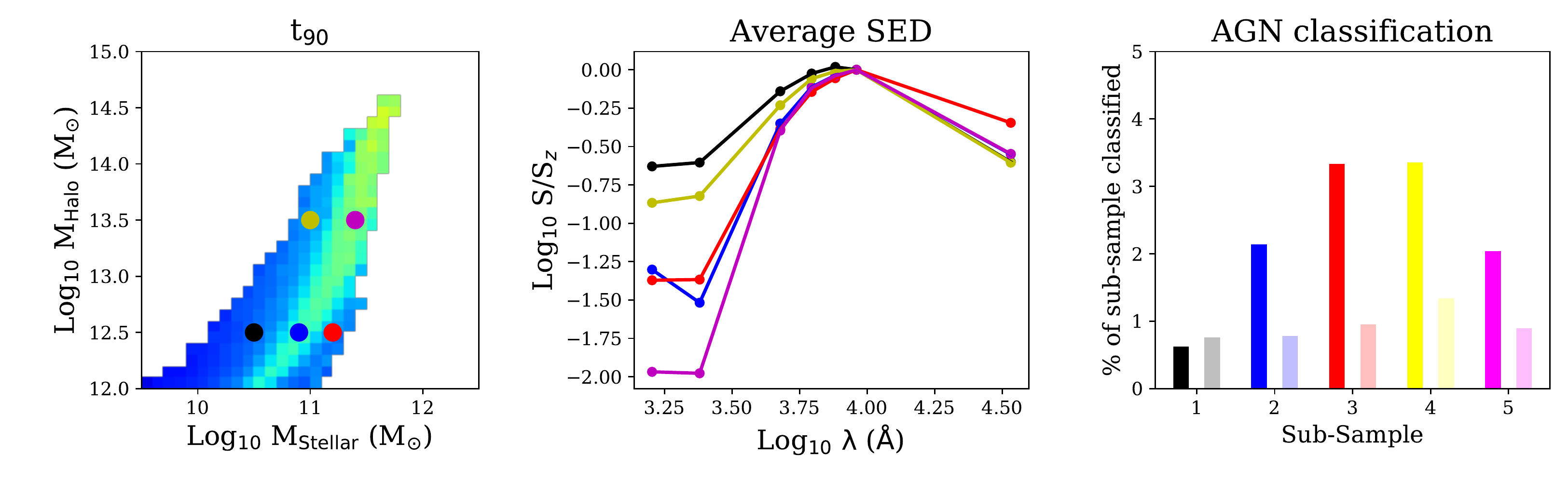}
		\caption{The left hand panel shows the location in stellar and halo mass space of the five selected sub-samples from our SDSS data. These are selected to occupy different parts of phase space in halo and stellar mass and have significantly different values of $t_{90}$. The central panel shows the average fluxes in each band for the five sub-populations, normalised to the {\it z}-band flux. The {\it W1} flux is noticeably higher for the high stellar-to-halo mass ratio sub-sample. The right hand panel shows the fraction of each sub-population selected as an AGN via either BPT \citep[from][solid bars]{2003MNRAS.346.1055K} or MIR selection \citep[from][shaded bars]{2018ApJS..234...23A}.}
		\label{fig:av_SED}
	\end{figure*}
	
	\begin{figure*}
		\centering
		\includegraphics[width=1.8\columnwidth]{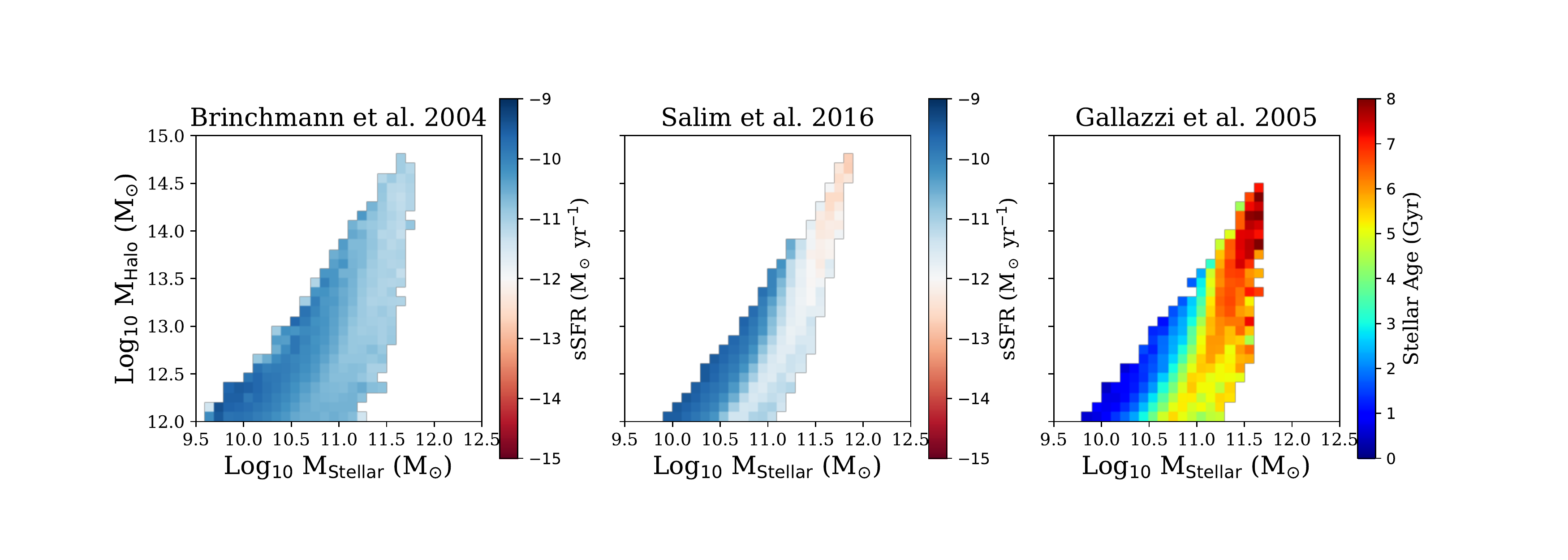}
		\caption{The left hand panel shows the sSFR as a function of halo and stellar mass as determined by \citet{2004MNRAS.351.1151B}, from using a mixture of emission lines and the 4000 \AA\ break. The central panel shows sSFR as a function of stellar and halo mass for our SDSS sample as estimated in the work of \citet{2016ApJS..227....2S}, which used SED fitting to determine stellar masses and SFRs. The right hand panel shows the luminosity weighted stellar ages as a function of stellar and halo mass from Gallazzi et al. (in prep). We see similar average trends as a function of stellar and halo mass as inferred from the SED fiting method of \citet{2016ApJ...824...45P}, albeit at somewhat different strengths due to differences in each SED fitting method, thereby showing the average trends are unlikely to be biased by our SED fitting method.}
		\label{fig:salim}
	\end{figure*}

	When we observe their growth from epoch to epoch in Figure~\ref{fig:grow_pac}, we see that they have a fairly constant and steady growth history. They do not grow as quickly at early epochs and as slowly at later epochs as the most massive galaxies, however not as quickly as the least massive galaxies at low redshift. They instead display an intermediate relative value at almost all epochs. We postulate that this steady growth is most likely driven by one of two scenarios. The first is steady gas accretion from the surrounding environment either via filaments or minor mergers, similar to the "cold mode" presented in \citet{2005MNRAS.363....2K}. The second is the cooling of recycled internal gas which has already been accreted at much earlier times such as postulated in \citet{2020arXiv200508995O}. Such galaxies have been studied, albeit in individual cases or small samples in \citet{2016ApJ...817..109O} and \citet{2019ApJS..243...14O}, where they also propose that these objects could be sustained by the cooling of already accreted gas.

	From Figure~\ref{fig:halo_comp}, by tracking our lowest stellar mass population as a function of halo mass (blue lines, mid left hand panel), we see that this sub-population is not as prominent in TNG300. We postulate that this rare population is likely to exist in TNG, albeit at much lower numbers. This is likely due to volume driven effects (the smaller the volume of the sample, the less likely we are to find peculiar objects statistically) as these galaxies are extreme and only make up 2 per cent of our observational sample.

	\subsection{Bias checking in the SED fits}
	\label{disc:sed_check}
	
	As the matches of an SED to the library of model star formation histories are dependent on the flux of the photometry itself, we decided to check the average SED shape for five sub-samples chosen to represent different populations of galaxies in halo and stellar mass, shown in the left panel of Figure~\ref{fig:av_SED}. These sub-samples were chosen to have bin widths of 0.2 dex in both halo and stellar mass and to contain a minimum of 50 galaxies. This was done in order to verify if any of the sub-populations had systematic biases that could affect our results. The mean photometric flux for each sub-sample in every photometric band was calculated and then normalised to the flux in the {\it z}-band, as this was the peak flux in four of the five sub-samples. This is shown in the central panel of Figure~\ref{fig:av_SED}. We see raised levels of UV flux in sub samples 1 and 4 (black and yellow line, Log$_{10}$ M$_{\mathrm{Stellar}} \sim$ 10.5 M$_\odot$ and Log$_{10}$ M$_{\mathrm{Halo}} \sim$ 12.5 M$_\odot$ and Log$_{10}$ M$_{\mathrm{Stellar}} \sim$ 11 M$_\odot$ and Log$_{10}$ M$_{\mathrm{Halo}} \sim$ 13.5 M$_\odot$ respectively) due to their raised level of star formation, as expected from Figure~\ref{fig:properties}. We also notice the raised {\it z} - W1 flux of sub-sample 3, our high stellar to halo mass ratio objects (red line, Log$_{10}$ M$_{\mathrm{Stellar}} \sim$ 11 M$_\odot$ and Log$_{10}$ M$_{\mathrm{Halo}} \sim$ 12.5 M$_\odot$), which is responsible for the high SFR estimates from the SED fitting.
	
	AGN emission can significantly contribute to the IR flux, especially in the MIR bands \citep{2016MNRAS.457.2703R}. Although type 1 AGN are removed from the catalogue of \citet{2016ApJ...824...45P}, type 2 AGN may still contribute to some of the IR emission. In order to check that this phenomenon was not biasing the SFR estimates and therefore the assembly histories in any of the sub-populations, with our high stellar to halo mass sub-population in particular, we cross matched the AGN catalogues of \citet{2003MNRAS.346.1055K} and \citet{2018ApJS..234...23A} with each sub-sample. We then calculated the percentage of each sub-sample that shows clear AGN activity, which can be seen in the right hand panel of Figure~\ref{fig:av_SED}. The solid bar for each sub-sample represents the percentage of galaxies found with AGN using the selection criteria of \citet{2003MNRAS.346.1055K} which are selected using the BPT selection criteria \citep*{1981PASP...93....5B}, and the shaded bar for each sub-sample represents those found in \citet{2018ApJS..234...23A}, which are identified using WISE MIR colour-colour selection. We see that the AGN fraction is consistently less than 5 per cent for all sub-samples in both selection criteria, and that sub-sample 3, our high stellar to halo mass objects, does not have a significantly higher fraction of AGN using either selection criteria, implying that it is unlikely that AGN emission biases the SFHs of this sub-sample.
	
	We also took data from the catalogues of \citet{2004MNRAS.351.1151B} and \citet{2016ApJS..227....2S}. Although these two catalogues do not contain mass assembly histories, we checked the stellar and halo mass distributions versus the sSFR, using the stellar mass and SFR estimates from each catalogue for consistency with the halo mass estimates from \citet{2017MNRAS.470.2982L}. If there is a systematic bias in the SED fitting method of \citet{2016ApJ...824...45P}, then these two catalogues should not reproduce the trends in the sSFR, stellar mass-halo mass plane we observe in Figure~\ref{fig:properties}. \citet{2004MNRAS.351.1151B} combine data from SDSS DR4 \citep{2006ApJS..162...38A} and a Bayesian approach to estimating SFRs and stellar masses from a mixture of emission lines and the 4000 \AA\ Balmer break. \citet{2016ApJS..227....2S}, alternatively, fit multi-wavelength photometry from {\it GALEX}, SDSS and {\it WISE} with the SED fitting code \textsc{cigale} \citep{2009A&A...507.1793N}. By cross matching these two catalogues with \citet{2017MNRAS.470.2982L}, we compared the sSFR as a function of stellar mass and halo mass, shown in Figure~\ref{fig:salim}. We note that both catalogues have higher upper limits on the sSFR than the catalogue of \citet{2016ApJ...824...45P}, meaning we searched for a replication of the general trends rather than absolute values.
	
	The left hand panel of Figure~\ref{fig:salim} shows the sSFR as a function of both stellar mass and halo mass from \citet{2004MNRAS.351.1151B}. We see subtle trends as a function of both stellar and halo mass, confirming the secondary dependencies we see in Figure~\ref{fig:properties}. We also see that the most massive galaxies (Log$_{10}$ M$_{\mathrm{Stellar}} \sim$ 11.5 M$_\odot$, Log$_{10}$ M$_{\mathrm{Halo}} \sim$ 14 M$_\odot$) have the lowest sSFRs and that the highest stellar-to-halo mass objects (Log$_{10}$ M$_{\mathrm{Stellar}} \sim$ 11 M$_\odot$ and Log$_{10}$ M$_{\mathrm{Halo}} \sim$ 12) have slightly higher sSFRs compared to objects of similar stellar mass but higher halo mass. These average trends are replicated, with more clear differences in various sub-populations, in the central panel from the catalogue of \citet{2016ApJS..227....2S}.
	
	We also use stellar mass and luminosity weighted stellar age estimations derived from SDSS spectra from Gallazzi et al. (in prep, method based on \citet{2005MNRAS.362...41G} and outlined in \citet{2019MNRAS.484.1702P}), in order to compare if spectroscopic techniques also reproduce our average trends in stellar and halo mass, this can be seen in the right hand panel of Figure~\ref{fig:salim}. We see similar average trends reproduced as in Figure~\ref{fig:properties}. We argue that the tests carried out in this section mean that the average trends found in previous sections are to a good degree of confidence, reliable.

	\subsection{Selection effects and sample matching}
	\label{disc:limits}
	
	As mentioned in Section~\ref{sim:matching}, to be able to make a meaningful comparison between the observed and simulated data sets, some level of matching needs to be applied. Techniques such as stellar or halo mass matching, star formation rate or colours are commonly used. We chose not to match in these factors as we would significantly reduce the size of our observational sample and we wanted to compare how well TNG300 reproduced the trends in stellar-to-halo mass with respect to the other variables, in order to best observe the trends produced by environmental and secular processes. We briefly consider other possibly matching techniques and their possible effects on the results.
	
	Due to cross-matching multiple group catalogues, which resulted in the loss of a significant number of galaxies, it is extremely difficult to calculate the volume of the observational sample, and beyond the scope of this work. We assume that TNG300 has a smaller volume than our observational sample due to the smaller sample size. This could mean that statistically we do not see the observed trend due to less extreme objects, however we argue that a ratio of 1:2 in the sample size difference would not be enough to hide the prominent behaviour of the highest stellar-to-halo mass objects and the differences in galaxies at intermediate stellar mass, which TNG300 should sample well. 
	
	Jackson et al. (submitted to MNRAS) applied a matching criterion between an observational and simulational sample of low redshift AGN. They used the Cartesian coordinates of each galaxy within the simulation box to calculate the distance from the centre of the simulation box. This was then combined with the luminosities of each galaxy to yield the fluxes, which could then be used to make a cut, mimicking the survey sensitivity limits. We repeated the same exercise here, from 100 different locations in the simulation box, in order to observe if a significant amount of objects were cut from the TNG300 sample according to the SDSS magnitude limit. We find that less than 1 per cent of all central galaxies above a stellar mass of 10$^9$ M$_{\odot}$ fall under the magnitude limit at any point in the simulation box, meaning the survey limit does not heavily bias our results. We tentatively conclude that selection effects from instrument sensitivity are unlikely to cause the differences we see between the observational and simulational trends.

	\section{Summary \& conclusions}
	\label{sec:conclusions}
	
	In this paper we have presented the stellar mass assembly of a sample of $\sim 90,000$ local, central galaxies in groups and clusters above a halo mass M$_{\mathrm{Halo}} > 10^{12}$ M$_\odot$ (M$_\odot \gtrsim 10^{9.5}$M$_\odot$). We matched SDSS group catalogues from \citet{2007ApJ...671..153Y}, \citet{2017MNRAS.470.2982L} and \citet{2005AJ....129.2562B} to obtain numerous galaxy properties such as halo mass, S\'{e}rsic index, number of satellite galaxies, effective radius and Malmquist bias. We then took estimates of the assembly times for 10, 50 and 90 per cent of the stellar mass and the relative growth at various epochs from \citet{2016ApJ...824...45P}. These estimates were obtained from fitting multi-wavelength photometry, measured within 2$R_e$ on average, from the UV to the NIR with SED models derived from realistic star formation histories generated via semi-analytic model of \citet{2007MNRAS.375....2D}. This allows a more in-depth investigation of the stellar mass build-up than single stellar ages alone. We then, for the first time, investigated the trends of these stellar assembly times simultaneously in the stellar versus halo mass plane in an attempt to separate secular and environmental processes and the effect they have on galaxy evolution.
	
	We find clear dependencies of all of the stellar assembly lookback times on stellar mass, whereby galaxies with higher stellar masses (at constant halo mass) have on average older lookback times, mirroring downsizing behaviour found by previous studies of galaxy assembly. We also find, however, secondary halo mass dependencies, whereby galaxies with higher halo mass at constant stellar mass have slightly younger assembly times. This could possibly due to either the larger potential of the more massive halo able to accrete and cool gas more efficiently or the ability to recycle old gas more efficiently. An exception to this behaviour is a sub sample of massive galaxies in relatively small haloes (Log$_{10}$ M$_{\mathrm{Stellar}} \sim$ 11 M$_\odot$ and Log$_{10}$ M$_{\mathrm{Halo}} \sim$ 12). By observing the growth rate throughout cosmic time as well as inspecting the average sSFR, S\'{e}rsic indices and the number of satellite galaxies, we conclude that these are likely massive, late-type, field galaxies. 
	
	We compared these results to the TNG300 simulation. We find that the simulations predict on average similar stellar assembly lookback times as either a function of stellar and halo mass, with trends generally within the scatter of the observational data. However, differences are found in secondary trends when both stellar and halo mass are simultaneously considered. We find these differences in behaviour in the assembly times as a function of stellar and halo mass most noticeably in intermediate stellar mass objects (Log$_{10}$ M$_{\mathrm{Stellar}} \sim$ 10.5 M$_\odot$ and Log$_{10}$ M$_{\mathrm{Halo}} \sim$ 13), whereby TNG300 predicts these should be some of the oldest objects in the sample, compared to the youngest in the SDSS sample. Discrepancies in the stellar assembly times on the stellar-halo mass plane manifest themselves also in differences in the quenched fractions between TNG300 and SDSS when these are evaluated simultaneously in bins of stellar and halo masses. As the kinetic mode black hole feedback displays similar behaviour as a function of stellar and halo mass, we tentatively link this to the difference in behaviour between the observations and simulations. A deeper investigation, beyond the scope of this work, would be required to confirm this link however.

	\section*{Acknowledgements}
	
	The authors would like to thank the anonymous referee for their thorough reading of this manuscript and their constructive comments and suggestions which have helped improve this work.
	
	Thomas Jackson is a fellow of the International Max Planck Research School for Astronomy and Cosmic Physics at the University of Heidelberg (IMPRS-HD). 
	
	The authors thank the IllustrisTNG team for making their data publically available.
	
	The authors thank Anna Gallazzi for the use of her SDSS based catalogue to compare age estimates from the SED fitting, with estimates obtained from stellar population fitting.
	
	Funding for the Sloan Digital Sky Survey IV has been provided by the Alfred P. Sloan Foundation, the U.S. Department of Energy Office of Science, and the Participating Institutions. SDSS-IV acknowledges
	support and resources from the Center for High-Performance Computing at
	the University of Utah. The SDSS web site is www.sdss.org.
	
	SDSS-IV is managed by the Astrophysical Research Consortium for the 
	Participating Institutions of the SDSS Collaboration including the 
	Brazilian Participation Group, the Carnegie Institution for Science, 
	Carnegie Mellon University, the Chilean Participation Group, the French Participation Group, Harvard-Smithsonian Center for Astrophysics, 
	Instituto de Astrof\'isica de Canarias, The Johns Hopkins University, Kavli Institute for the Physics and Mathematics of the Universe (IPMU) / 
	University of Tokyo, the Korean Participation Group, Lawrence Berkeley National Laboratory, 
	Leibniz Institut f\"ur Astrophysik Potsdam (AIP),  
	Max-Planck-Institut f\"ur Astronomie (MPIA Heidelberg), 
	Max-Planck-Institut f\"ur Astrophysik (MPA Garching), 
	Max-Planck-Institut f\"ur Extraterrestrische Physik (MPE), 
	National Astronomical Observatories of China, New Mexico State University, 
	New York University, University of Notre Dame, 
	Observat\'ario Nacional / MCTI, The Ohio State University, 
	Pennsylvania State University, Shanghai Astronomical Observatory, 
	United Kingdom Participation Group,
	Universidad Nacional Aut\'onoma de M\'exico, University of Arizona, 
	University of Colorado Boulder, University of Oxford, University of Portsmouth, 
	University of Utah, University of Virginia, University of Washington, University of Wisconsin, 
	Vanderbilt University, and Yale University.
	
	This publication makes use of data products from the Wide-field Infrared Survey Explorer, which is a joint project of the University of California, Los Angeles, and the Jet Propulsion Laboratory/California Institute of Technology, funded by the National Aeronautics and Space Administration.
	
	{\it GALEX} (Galaxy Evolution Explorer) is a NASA Small Explorer, launched in 2003 April. We gratefully acknowledge NASA\textquotesingle s support for the construction, operation, and science analysis for the {\it GALEX} mission, developed incooperation with the Centre National d\textquotesingle Etudes Spatiales of France and the Korean Ministry of Science and Technology.
	
	The flagship simulations of the IllustrisTNG project used in this work have been run on the HazelHen Cray XC40-system at the High Performance Computing Center Stuttgart as part of project GCS-ILLU of the Gauss centres for Super-computing(GCS). Ancillary and test runs of the project were also run on the Stampede supercomputer at TACC/XSEDE (allocation AST140063), at the Hydra and Draco supercomputers at the Max Planck Computing and Data Facility, and on the MIT/Harvard computing facilities supported by FAS and MIT MKI.

	\section{Data availability}
	
	The data underlying this article are available from the public sources in the links or references given in the article (or references therein). Estimations from the article are available on request.
	
	
	
	
	\bibliographystyle{mnras}
	\bibliography{galaxy_evolution} 

\begin{thebibliography}{}
\makeatletter
\relax
\def\mn@urlcharsother{\let\do\@makeother \do\$\do\&\do\#\do\^\do\_\do\%\do\~}
\def\mn@doi{\begingroup\mn@urlcharsother \@ifnextchar [ {\mn@doi@}
  {\mn@doi@[]}}
\def\mn@doi@[#1]#2{\def\@tempa{#1}\ifx\@tempa\@empty \href
  {http://dx.doi.org/#2} {doi:#2}\else \href {http://dx.doi.org/#2} {#1}\fi
  \endgroup}
\def\mn@eprint#1#2{\mn@eprint@#1:#2::\@nil}
\def\mn@eprint@arXiv#1{\href {http://arxiv.org/abs/#1} {{\tt arXiv:#1}}}
\def\mn@eprint@dblp#1{\href {http://dblp.uni-trier.de/rec/bibtex/#1.xml}
  {dblp:#1}}
\def\mn@eprint@#1:#2:#3:#4\@nil{\def\@tempa {#1}\def\@tempb {#2}\def\@tempc
  {#3}\ifx \@tempc \@empty \let \@tempc \@tempb \let \@tempb \@tempa \fi \ifx
  \@tempb \@empty \def\@tempb {arXiv}\fi \@ifundefined
  {mn@eprint@\@tempb}{\@tempb:\@tempc}{\expandafter \expandafter \csname
  mn@eprint@\@tempb\endcsname \expandafter{\@tempc}}}

\bibitem[\protect\citeauthoryear{{Abazajian} et~al.,}{{Abazajian}
  et~al.}{2009}]{2009ApJS..182..543A}
{Abazajian} K.~N.,  et~al., 2009, \mn@doi [\apjs]
  {10.1088/0067-0049/182/2/543}, \href
  {https://ui.adsabs.harvard.edu/abs/2009ApJS..182..543A} {182, 543}

\bibitem[\protect\citeauthoryear{{Adelman-McCarthy} et~al.,}{{Adelman-McCarthy}
  et~al.}{2006}]{2006ApJS..162...38A}
{Adelman-McCarthy} J.~K.,  et~al., 2006, \mn@doi [\apjs] {10.1086/497917},
  \href {https://ui.adsabs.harvard.edu/abs/2006ApJS..162...38A} {162, 38}

\bibitem[\protect\citeauthoryear{{Ahn} et~al.,}{{Ahn}
  et~al.}{2014}]{2014ApJS..211...17A}
{Ahn} C.~P.,  et~al., 2014, \mn@doi [\apjs] {10.1088/0067-0049/211/2/17}, \href
  {https://ui.adsabs.harvard.edu/abs/2014ApJS..211...17A} {211, 17}

\bibitem[\protect\citeauthoryear{{Albareti} et~al.,}{{Albareti}
  et~al.}{2017}]{2017ApJS..233...25A}
{Albareti} F.~D.,  et~al., 2017, \mn@doi [\apjs] {10.3847/1538-4365/aa8992},
  \href {https://ui.adsabs.harvard.edu/abs/2017ApJS..233...25A} {233, 25}

\bibitem[\protect\citeauthoryear{{Assef}, {Stern}, {Noirot}, {Jun}, {Cutri}  \&
  {Eisenhardt}}{{Assef} et~al.}{2018}]{2018ApJS..234...23A}
{Assef} R.~J.,  {Stern} D.,  {Noirot} G.,  {Jun} H.~D.,  {Cutri} R.~M.,
  {Eisenhardt} P.~R.~M.,  2018, \mn@doi [\apjs] {10.3847/1538-4365/aaa00a},
  \href {https://ui.adsabs.harvard.edu/abs/2018ApJS..234...23A} {234, 23}

\bibitem[\protect\citeauthoryear{{Baldwin}, {Phillips}  \&
  {Terlevich}}{{Baldwin} et~al.}{1981}]{1981PASP...93....5B}
{Baldwin} J.~A.,  {Phillips} M.~M.,   {Terlevich} R.,  1981, \mn@doi [\pasp]
  {10.1086/130766}, \href
  {https://ui.adsabs.harvard.edu/abs/1981PASP...93....5B} {93, 5}

\bibitem[\protect\citeauthoryear{{Bezanson}, {van Dokkum}, {Tal}, {Marchesini},
  {Kriek}, {Franx}  \& {Coppi}}{{Bezanson} et~al.}{2009}]{2009ApJ...697.1290B}
{Bezanson} R.,  {van Dokkum} P.~G.,  {Tal} T.,  {Marchesini} D.,  {Kriek} M.,
  {Franx} M.,   {Coppi} P.,  2009, \mn@doi [\apj]
  {10.1088/0004-637X/697/2/1290}, \href
  {http://adsabs.harvard.edu/abs/2009ApJ...697.1290B} {697, 1290}

\bibitem[\protect\citeauthoryear{{Blanton} et~al.,}{{Blanton}
  et~al.}{2005}]{2005AJ....129.2562B}
{Blanton} M.~R.,  et~al., 2005, \mn@doi [\aj] {10.1086/429803}, \href
  {http://adsabs.harvard.edu/abs/2005AJ....129.2562B} {129, 2562}

\bibitem[\protect\citeauthoryear{{Bower}, {Benson}, {Malbon}, {Helly}, {Frenk},
  {Baugh}, {Cole}  \& {Lacey}}{{Bower} et~al.}{2006}]{2006MNRAS.370..645B}
{Bower} R.~G.,  {Benson} A.~J.,  {Malbon} R.,  {Helly} J.~C.,  {Frenk} C.~S.,
  {Baugh} C.~M.,  {Cole} S.,   {Lacey} C.~G.,  2006, \mn@doi [\mnras]
  {10.1111/j.1365-2966.2006.10519.x}, \href
  {http://adsabs.harvard.edu/abs/2006MNRAS.370..645B} {370, 645}

\bibitem[\protect\citeauthoryear{{Brinchmann}, {Charlot}, {White}, {Tremonti},
  {Kauffmann}, {Heckman}  \& {Brinkmann}}{{Brinchmann}
  et~al.}{2004}]{2004MNRAS.351.1151B}
{Brinchmann} J.,  {Charlot} S.,  {White} S.~D.~M.,  {Tremonti} C.,  {Kauffmann}
  G.,  {Heckman} T.,   {Brinkmann} J.,  2004, \mn@doi [\mnras]
  {10.1111/j.1365-2966.2004.07881.x}, \href
  {http://adsabs.harvard.edu/abs/2004MNRAS.351.1151B} {351, 1151}

\bibitem[\protect\citeauthoryear{{Bruzual} \& {Charlot}}{{Bruzual} \&
  {Charlot}}{2003}]{2003MNRAS.344.1000B}
{Bruzual} G.,  {Charlot} S.,  2003, \mn@doi [\mnras]
  {10.1046/j.1365-8711.2003.06897.x}, \href
  {https://ui.adsabs.harvard.edu/abs/2003MNRAS.344.1000B} {344, 1000}

\bibitem[\protect\citeauthoryear{{Charlot} \& {Fall}}{{Charlot} \&
  {Fall}}{2000}]{2000ApJ...539..718C}
{Charlot} S.,  {Fall} S.~M.,  2000, \mn@doi [\apj] {10.1086/309250}, \href
  {http://adsabs.harvard.edu/abs/2000ApJ...539..718C} {539, 718}

\bibitem[\protect\citeauthoryear{{Conroy}, {Gunn}  \& {White}}{{Conroy}
  et~al.}{2009}]{2009ApJ...699..486C}
{Conroy} C.,  {Gunn} J.~E.,   {White} M.,  2009, \mn@doi [\apj]
  {10.1088/0004-637X/699/1/486}, \href
  {https://ui.adsabs.harvard.edu/abs/2009ApJ...699..486C} {699, 486}

\bibitem[\protect\citeauthoryear{{Cowie}, {Songaila}, {Hu}  \& {Cohen}}{{Cowie}
  et~al.}{1996}]{1996AJ....112..839C}
{Cowie} L.~L.,  {Songaila} A.,  {Hu} E.~M.,   {Cohen} J.~G.,  1996, \mn@doi
  [\aj] {10.1086/118058}, \href
  {https://ui.adsabs.harvard.edu/abs/1996AJ....112..839C} {112, 839}

\bibitem[\protect\citeauthoryear{{Crain} et~al.,}{{Crain}
  et~al.}{2015}]{2015MNRAS.450.1937C}
{Crain} R.~A.,  et~al., 2015, \mn@doi [\mnras] {10.1093/mnras/stv725}, \href
  {https://ui.adsabs.harvard.edu/abs/2015MNRAS.450.1937C} {450, 1937}

\bibitem[\protect\citeauthoryear{{Daddi} et~al.,}{{Daddi}
  et~al.}{2005}]{2005ApJ...626..680D}
{Daddi} E.,  et~al., 2005, \mn@doi [\apj] {10.1086/430104}, \href
  {http://adsabs.harvard.edu/abs/2005ApJ...626..680D} {626, 680}

\bibitem[\protect\citeauthoryear{{De Lucia}}{{De
  Lucia}}{2011}]{2011ASSP...27..203D}
{De Lucia} G.,  2011, \mn@doi [Astrophysics and Space Science Proceedings]
  {10.1007/978-3-642-20285-8_41}, \href
  {https://ui.adsabs.harvard.edu/abs/2011ASSP...27..203D} {27, 203}

\bibitem[\protect\citeauthoryear{{De Lucia} \& {Blaizot}}{{De Lucia} \&
  {Blaizot}}{2007}]{2007MNRAS.375....2D}
{De Lucia} G.,  {Blaizot} J.,  2007, \mn@doi [\mnras]
  {10.1111/j.1365-2966.2006.11287.x}, \href
  {http://adsabs.harvard.edu/abs/2007MNRAS.375....2D} {375, 2}

\bibitem[\protect\citeauthoryear{{Donnari} et~al.,}{{Donnari}
  et~al.}{2019}]{2019MNRAS.485.4817D}
{Donnari} M.,  et~al., 2019, \mn@doi [\mnras] {10.1093/mnras/stz712}, \href
  {https://ui.adsabs.harvard.edu/abs/2019MNRAS.485.4817D} {485, 4817}

\bibitem[\protect\citeauthoryear{{Dubois} et~al.,}{{Dubois}
  et~al.}{2014}]{2014MNRAS.444.1453D}
{Dubois} Y.,  et~al., 2014, \mn@doi [\mnras] {10.1093/mnras/stu1227}, \href
  {https://ui.adsabs.harvard.edu/abs/2014MNRAS.444.1453D} {444, 1453}

\bibitem[\protect\citeauthoryear{{Ferland}, {Korista}, {Verner}, {Ferguson},
  {Kingdon}  \& {Verner}}{{Ferland} et~al.}{1998}]{1998PASP..110..761F}
{Ferland} G.~J.,  {Korista} K.~T.,  {Verner} D.~A.,  {Ferguson} J.~W.,
  {Kingdon} J.~B.,   {Verner} E.~M.,  1998, \mn@doi [\pasp] {10.1086/316190},
  \href {http://adsabs.harvard.edu/abs/1998PASP..110..761F} {110, 761}

\bibitem[\protect\citeauthoryear{{Ferland} et~al.,}{{Ferland}
  et~al.}{2017}]{2017RMxAA..53..385F}
{Ferland} G.~J.,  et~al., 2017, \rmxaa, \href
  {http://adsabs.harvard.edu/abs/2017RMxAA..53..385F} {53, 385}

\bibitem[\protect\citeauthoryear{{Fitzpatrick}}{{Fitzpatrick}}{1999}]{1999PASP..111...63F}
{Fitzpatrick} E.~L.,  1999, \mn@doi [\pasp] {10.1086/316293}, \href
  {https://ui.adsabs.harvard.edu/abs/1999PASP..111...63F} {111, 63}

\bibitem[\protect\citeauthoryear{{Furlong} et~al.,}{{Furlong}
  et~al.}{2015}]{2015MNRAS.450.4486F}
{Furlong} M.,  et~al., 2015, \mn@doi [\mnras] {10.1093/mnras/stv852}, \href
  {https://ui.adsabs.harvard.edu/abs/2015MNRAS.450.4486F} {450, 4486}

\bibitem[\protect\citeauthoryear{{Gallazzi}, {Charlot}, {Brinchmann}, {White}
  \& {Tremonti}}{{Gallazzi} et~al.}{2005}]{2005MNRAS.362...41G}
{Gallazzi} A.,  {Charlot} S.,  {Brinchmann} J.,  {White} S. D.~M.,   {Tremonti}
  C.~A.,  2005, \mn@doi [\mnras] {10.1111/j.1365-2966.2005.09321.x}, \href
  {https://ui.adsabs.harvard.edu/abs/2005MNRAS.362...41G} {362, 41}

\bibitem[\protect\citeauthoryear{{Genel} et~al.,}{{Genel}
  et~al.}{2014}]{2014MNRAS.445..175G}
{Genel} S.,  et~al., 2014, \mn@doi [\mnras] {10.1093/mnras/stu1654}, \href
  {http://adsabs.harvard.edu/abs/2014MNRAS.445..175G} {445, 175}

\bibitem[\protect\citeauthoryear{{Genel} et~al.,}{{Genel}
  et~al.}{2018}]{2018MNRAS.474.3976G}
{Genel} S.,  et~al., 2018, \mn@doi [\mnras] {10.1093/mnras/stx3078}, \href
  {http://adsabs.harvard.edu/abs/2018MNRAS.474.3976G} {474, 3976}

\bibitem[\protect\citeauthoryear{{Jackson}, {Martin}, {Kaviraj}, {Laigle},
  {Devriendt}, {Dubois}  \& {Pichon}}{{Jackson}
  et~al.}{2020}]{2020MNRAS.494.5568J}
{Jackson} R.~A.,  {Martin} G.,  {Kaviraj} S.,  {Laigle} C.,  {Devriendt}
  J.~E.~G.,  {Dubois} Y.,   {Pichon} C.,  2020, \mn@doi [\mnras]
  {10.1093/mnras/staa970}, \href
  {https://ui.adsabs.harvard.edu/abs/2020MNRAS.494.5568J} {494, 5568}

\bibitem[\protect\citeauthoryear{{Katz}, {Keres}, {Dave}  \& {Weinberg}}{{Katz}
  et~al.}{2003}]{2003ASSL..281..185K}
{Katz} N.,  {Keres} D.,  {Dave} R.,   {Weinberg} D.~H.,  2003, {How Do Galaxies
  Get Their Gas?}.
p.~185, \mn@doi{10.1007/978-94-010-0115-1_34}

\bibitem[\protect\citeauthoryear{{Kauffmann} et~al.,}{{Kauffmann}
  et~al.}{2003}]{2003MNRAS.346.1055K}
{Kauffmann} G.,  et~al., 2003, \mn@doi [\mnras]
  {10.1111/j.1365-2966.2003.07154.x}, \href
  {https://ui.adsabs.harvard.edu/abs/2003MNRAS.346.1055K} {346, 1055}

\bibitem[\protect\citeauthoryear{{Kennicutt}}{{Kennicutt}}{1998}]{1998ApJ...498..541K}
{Kennicutt} Robert~C. J.,  1998, \mn@doi [\apj] {10.1086/305588}, \href
  {https://ui.adsabs.harvard.edu/abs/1998ApJ...498..541K} {498, 541}

\bibitem[\protect\citeauthoryear{{Kere{\v{s}}}, {Katz}, {Weinberg}  \&
  {Dav{\'e}}}{{Kere{\v{s}}} et~al.}{2005}]{2005MNRAS.363....2K}
{Kere{\v{s}}} D.,  {Katz} N.,  {Weinberg} D.~H.,   {Dav{\'e}} R.,  2005,
  \mn@doi [\mnras] {10.1111/j.1365-2966.2005.09451.x}, \href
  {https://ui.adsabs.harvard.edu/abs/2005MNRAS.363....2K} {363, 2}

\bibitem[\protect\citeauthoryear{{La Barbera}, {Ferreras}, {de Carvalho},
  {Bruzual}, {Charlot}, {Pasquali}  \& {Merlin}}{{La Barbera}
  et~al.}{2012}]{2012MNRAS.426.2300L}
{La Barbera} F.,  {Ferreras} I.,  {de Carvalho} R.~R.,  {Bruzual} G.,
  {Charlot} S.,  {Pasquali} A.,   {Merlin} E.,  2012, \mn@doi [\mnras]
  {10.1111/j.1365-2966.2012.21848.x}, \href
  {https://ui.adsabs.harvard.edu/abs/2012MNRAS.426.2300L} {426, 2300}

\bibitem[\protect\citeauthoryear{{La Barbera}, {Pasquali}, {Ferreras},
  {Gallazzi}, {de Carvalho}  \& {de la Rosa}}{{La Barbera}
  et~al.}{2014}]{2014MNRAS.445.1977L}
{La Barbera} F.,  {Pasquali} A.,  {Ferreras} I.,  {Gallazzi} A.,  {de Carvalho}
  R.~R.,   {de la Rosa} I.~G.,  2014, \mn@doi [\mnras] {10.1093/mnras/stu1626},
  \href {https://ui.adsabs.harvard.edu/abs/2014MNRAS.445.1977L} {445, 1977}

\bibitem[\protect\citeauthoryear{{La Barbera} et~al.,}{{La Barbera}
  et~al.}{2019}]{2019MNRAS.489.4090L}
{La Barbera} F.,  et~al., 2019, \mn@doi [\mnras] {10.1093/mnras/stz2192}, \href
  {https://ui.adsabs.harvard.edu/abs/2019MNRAS.489.4090L} {489, 4090}

\bibitem[\protect\citeauthoryear{{Lim}, {Mo}, {Lu}, {Wang}  \& {Yang}}{{Lim}
  et~al.}{2017}]{2017MNRAS.470.2982L}
{Lim} S.~H.,  {Mo} H.~J.,  {Lu} Y.,  {Wang} H.,   {Yang} X.,  2017, \mn@doi
  [\mnras] {10.1093/mnras/stx1462}, \href
  {http://adsabs.harvard.edu/abs/2017MNRAS.470.2982L} {470, 2982}

\bibitem[\protect\citeauthoryear{{Lu}, {Mo}, {Lu}, {Katz}, {Weinberg}, {van den
  Bosch}  \& {Yang}}{{Lu} et~al.}{2015}]{2015MNRAS.450.1604L}
{Lu} Z.,  {Mo} H.~J.,  {Lu} Y.,  {Katz} N.,  {Weinberg} M.~D.,  {van den Bosch}
  F.~C.,   {Yang} X.,  2015, \mn@doi [\mnras] {10.1093/mnras/stv667}, \href
  {https://ui.adsabs.harvard.edu/abs/2015MNRAS.450.1604L} {450, 1604}

\bibitem[\protect\citeauthoryear{{Maraston}}{{Maraston}}{2005}]{2005MNRAS.362..799M}
{Maraston} C.,  2005, \mn@doi [\mnras] {10.1111/j.1365-2966.2005.09270.x},
  \href {https://ui.adsabs.harvard.edu/abs/2005MNRAS.362..799M} {362, 799}

\bibitem[\protect\citeauthoryear{{Marinacci} et~al.,}{{Marinacci}
  et~al.}{2018}]{2018MNRAS.480.5113M}
{Marinacci} F.,  et~al., 2018, \mn@doi [\mnras] {10.1093/mnras/sty2206}, \href
  {http://adsabs.harvard.edu/abs/2018MNRAS.480.5113M} {480, 5113}

\bibitem[\protect\citeauthoryear{{Mart{\'\i}n-Navarro}, {La Barbera},
  {Vazdekis}, {Falc{\'o}n-Barroso}  \& {Ferreras}}{{Mart{\'\i}n-Navarro}
  et~al.}{2015}]{2015MNRAS.447.1033M}
{Mart{\'\i}n-Navarro} I.,  {La Barbera} F.,  {Vazdekis} A.,
  {Falc{\'o}n-Barroso} J.,   {Ferreras} I.,  2015, \mn@doi [\mnras]
  {10.1093/mnras/stu2480}, \href
  {https://ui.adsabs.harvard.edu/abs/2015MNRAS.447.1033M} {447, 1033}

\bibitem[\protect\citeauthoryear{{McAlpine}, {Bower}, {Harrison}, {Crain},
  {Schaller}, {Schaye}  \& {Theuns}}{{McAlpine}
  et~al.}{2017}]{2017MNRAS.468.3395M}
{McAlpine} S.,  {Bower} R.~G.,  {Harrison} C.~M.,  {Crain} R.~A.,  {Schaller}
  M.,  {Schaye} J.,   {Theuns} T.,  2017, \mn@doi [\mnras]
  {10.1093/mnras/stx658}, \href
  {http://adsabs.harvard.edu/abs/2017MNRAS.468.3395M} {468, 3395}

\bibitem[\protect\citeauthoryear{{Naab}, {Johansson}  \& {Ostriker}}{{Naab}
  et~al.}{2009}]{2009ApJ...699L.178N}
{Naab} T.,  {Johansson} P.~H.,   {Ostriker} J.~P.,  2009, \mn@doi [\apjl]
  {10.1088/0004-637X/699/2/L178}, \href
  {http://adsabs.harvard.edu/abs/2009ApJ...699L.178N} {699, L178}

\bibitem[\protect\citeauthoryear{{Naiman} et~al.,}{{Naiman}
  et~al.}{2018}]{2018MNRAS.477.1206N}
{Naiman} J.~P.,  et~al., 2018, \mn@doi [\mnras] {10.1093/mnras/sty618}, \href
  {http://adsabs.harvard.edu/abs/2018MNRAS.477.1206N} {477, 1206}

\bibitem[\protect\citeauthoryear{{Navarro}, {Frenk}  \& {White}}{{Navarro}
  et~al.}{1997}]{1997ApJ...490..493N}
{Navarro} J.~F.,  {Frenk} C.~S.,   {White} S. D.~M.,  1997, \mn@doi [\apj]
  {10.1086/304888}, \href
  {https://ui.adsabs.harvard.edu/abs/1997ApJ...490..493N} {490, 493}

\bibitem[\protect\citeauthoryear{{Nelson} et~al.,}{{Nelson}
  et~al.}{2015}]{2015A&C....13...12N}
{Nelson} D.,  et~al., 2015, \mn@doi [Astronomy and Computing]
  {10.1016/j.ascom.2015.09.003}, \href
  {http://adsabs.harvard.edu/abs/2015A%26C....13...12N} {13, 12}

\bibitem[\protect\citeauthoryear{{Nelson} et~al.,}{{Nelson}
  et~al.}{2018}]{2018MNRAS.475..624N}
{Nelson} D.,  et~al., 2018, \mn@doi [\mnras] {10.1093/mnras/stx3040}, \href
  {http://adsabs.harvard.edu/abs/2018MNRAS.475..624N} {475, 624}

\bibitem[\protect\citeauthoryear{{Nelson} et~al.,}{{Nelson}
  et~al.}{2019}]{2019ComAC...6....2N}
{Nelson} D.,  et~al., 2019, \mn@doi [Computational Astrophysics and Cosmology]
  {10.1186/s40668-019-0028-x}, \href
  {https://ui.adsabs.harvard.edu/abs/2019ComAC...6....2N} {6, 2}

\bibitem[\protect\citeauthoryear{{Noll}, {Burgarella}, {Giovannoli}, {Buat},
  {Marcillac}  \& {Mu{\~n}oz-Mateos}}{{Noll}
  et~al.}{2009}]{2009A&A...507.1793N}
{Noll} S.,  {Burgarella} D.,  {Giovannoli} E.,  {Buat} V.,  {Marcillac} D.,
  {Mu{\~n}oz-Mateos} J.~C.,  2009, \mn@doi [\aap]
  {10.1051/0004-6361/200912497}, \href
  {https://ui.adsabs.harvard.edu/abs/2009A&A...507.1793N} {507, 1793}

\bibitem[\protect\citeauthoryear{{O'Donnell}, {Behroozi}  \&
  {More}}{{O'Donnell} et~al.}{2020}]{2020arXiv200508995O}
{O'Donnell} C.,  {Behroozi} P.,   {More} S.,  2020, arXiv e-prints, \href
  {https://ui.adsabs.harvard.edu/abs/2020arXiv200508995O} {p. arXiv:2005.08995}

\bibitem[\protect\citeauthoryear{{Ogle}, {Lanz}, {Nader}  \& {Helou}}{{Ogle}
  et~al.}{2016}]{2016ApJ...817..109O}
{Ogle} P.~M.,  {Lanz} L.,  {Nader} C.,   {Helou} G.,  2016, \mn@doi [\apj]
  {10.3847/0004-637X/817/2/109}, \href
  {http://adsabs.harvard.edu/abs/2016ApJ...817..109O} {817, 109}

\bibitem[\protect\citeauthoryear{{Ogle}, {Lanz}, {Appleton}, {Helou}  \&
  {Mazzarella}}{{Ogle} et~al.}{2019}]{2019ApJS..243...14O}
{Ogle} P.~M.,  {Lanz} L.,  {Appleton} P.~N.,  {Helou} G.,   {Mazzarella} J.,
  2019, \mn@doi [\apjs] {10.3847/1538-4365/ab21c3}, \href
  {https://ui.adsabs.harvard.edu/abs/2019ApJS..243...14O} {243, 14}

\bibitem[\protect\citeauthoryear{{Oh}, {Sarzi}, {Schawinski}  \& {Yi}}{{Oh}
  et~al.}{2011}]{2011ApJS..195...13O}
{Oh} K.,  {Sarzi} M.,  {Schawinski} K.,   {Yi} S.~K.,  2011, \mn@doi [\apjs]
  {10.1088/0067-0049/195/2/13}, \href
  {http://adsabs.harvard.edu/abs/2011ApJS..195...13O} {195, 13}

\bibitem[\protect\citeauthoryear{{Pacifici}, {Charlot}, {Blaizot}  \&
  {Brinchmann}}{{Pacifici} et~al.}{2012}]{2012MNRAS.421.2002P}
{Pacifici} C.,  {Charlot} S.,  {Blaizot} J.,   {Brinchmann} J.,  2012, \mn@doi
  [\mnras] {10.1111/j.1365-2966.2012.20431.x}, \href
  {http://adsabs.harvard.edu/abs/2012MNRAS.421.2002P} {421, 2002}

\bibitem[\protect\citeauthoryear{{Pacifici}, {Oh}, {Oh}, {Lee}  \&
  {Yi}}{{Pacifici} et~al.}{2016}]{2016ApJ...824...45P}
{Pacifici} C.,  {Oh} S.,  {Oh} K.,  {Lee} J.,   {Yi} S.~K.,  2016, \mn@doi
  [\apj] {10.3847/0004-637X/824/1/45}, \href
  {http://adsabs.harvard.edu/abs/2016ApJ...824...45P} {824, 45}

\bibitem[\protect\citeauthoryear{{Pasquali}, {van den Bosch}, {Mo}, {Yang}  \&
  {Somerville}}{{Pasquali} et~al.}{2009}]{2009MNRAS.394...38P}
{Pasquali} A.,  {van den Bosch} F.~C.,  {Mo} H.~J.,  {Yang} X.,   {Somerville}
  R.,  2009, \mn@doi [\mnras] {10.1111/j.1365-2966.2008.14233.x}, \href
  {https://ui.adsabs.harvard.edu/abs/2009MNRAS.394...38P} {394, 38}

\bibitem[\protect\citeauthoryear{{Pasquali}, {Gallazzi}, {Fontanot}, {van den
  Bosch}, {De Lucia}, {Mo}  \& {Yang}}{{Pasquali}
  et~al.}{2010}]{2010MNRAS.407..937P}
{Pasquali} A.,  {Gallazzi} A.,  {Fontanot} F.,  {van den Bosch} F.~C.,  {De
  Lucia} G.,  {Mo} H.~J.,   {Yang} X.,  2010, \mn@doi [\mnras]
  {10.1111/j.1365-2966.2010.17074.x}, \href
  {https://ui.adsabs.harvard.edu/abs/2010MNRAS.407..937P} {407, 937}

\bibitem[\protect\citeauthoryear{{Pasquali}, {Smith}, {Gallazzi}, {De Lucia},
  {Zibetti}, {Hirschmann}  \& {Yi}}{{Pasquali}
  et~al.}{2019}]{2019MNRAS.484.1702P}
{Pasquali} A.,  {Smith} R.,  {Gallazzi} A.,  {De Lucia} G.,  {Zibetti} S.,
  {Hirschmann} M.,   {Yi} S.~K.,  2019, \mn@doi [\mnras]
  {10.1093/mnras/sty3530}, \href
  {https://ui.adsabs.harvard.edu/abs/2019MNRAS.484.1702P} {484, 1702}

\bibitem[\protect\citeauthoryear{{Pillepich} et~al.,}{{Pillepich}
  et~al.}{2018a}]{2018MNRAS.473.4077P}
{Pillepich} A.,  et~al., 2018a, \mn@doi [\mnras] {10.1093/mnras/stx2656}, \href
  {http://adsabs.harvard.edu/abs/2018MNRAS.473.4077P} {473, 4077}

\bibitem[\protect\citeauthoryear{{Pillepich} et~al.,}{{Pillepich}
  et~al.}{2018b}]{2018MNRAS.475..648P}
{Pillepich} A.,  et~al., 2018b, \mn@doi [\mnras] {10.1093/mnras/stx3112}, \href
  {http://adsabs.harvard.edu/abs/2018MNRAS.475..648P} {475, 648}

\bibitem[\protect\citeauthoryear{{Pillepich} et~al.,}{{Pillepich}
  et~al.}{2019}]{2019MNRAS.490.3196P}
{Pillepich} A.,  et~al., 2019, \mn@doi [\mnras] {10.1093/mnras/stz2338}, \href
  {https://ui.adsabs.harvard.edu/abs/2019MNRAS.490.3196P} {490, 3196}

\bibitem[\protect\citeauthoryear{{Planck Collaboration} et~al.,}{{Planck
  Collaboration} et~al.}{2016}]{2016A&A...594A..13P}
{Planck Collaboration} et~al., 2016, \mn@doi [\aap]
  {10.1051/0004-6361/201525830}, \href
  {https://ui.adsabs.harvard.edu/abs/2016A&A...594A..13P} {594, A13}

\bibitem[\protect\citeauthoryear{{Renzini}}{{Renzini}}{2006}]{2006ARA&A..44..141R}
{Renzini} A.,  2006, \mn@doi [\araa] {10.1146/annurev.astro.44.051905.092450},
  \href {https://ui.adsabs.harvard.edu/abs/2006ARA&A..44..141R} {44, 141}

\bibitem[\protect\citeauthoryear{{Rosario}, {Mendel}, {Ellison}, {Lutz}  \&
  {Trump}}{{Rosario} et~al.}{2016}]{2016MNRAS.457.2703R}
{Rosario} D.~J.,  {Mendel} J.~T.,  {Ellison} S.~L.,  {Lutz} D.,   {Trump}
  J.~R.,  2016, \mn@doi [\mnras] {10.1093/mnras/stw096}, \href
  {http://adsabs.harvard.edu/abs/2016MNRAS.457.2703R} {457, 2703}

\bibitem[\protect\citeauthoryear{{Salim} et~al.,}{{Salim}
  et~al.}{2016}]{2016ApJS..227....2S}
{Salim} S.,  et~al., 2016, \mn@doi [\apjs] {10.3847/0067-0049/227/1/2}, \href
  {http://adsabs.harvard.edu/abs/2016ApJS..227....2S} {227, 2}

\bibitem[\protect\citeauthoryear{{Sanders} \& {Mirabel}}{{Sanders} \&
  {Mirabel}}{1996}]{1996ARA&A..34..749S}
{Sanders} D.~B.,  {Mirabel} I.~F.,  1996, \mn@doi [\araa]
  {10.1146/annurev.astro.34.1.749}, \href
  {https://ui.adsabs.harvard.edu/abs/1996ARA&A..34..749S} {34, 749}

\bibitem[\protect\citeauthoryear{{Schaye} et~al.,}{{Schaye}
  et~al.}{2015}]{2015MNRAS.446..521S}
{Schaye} J.,  et~al., 2015, \mn@doi [\mnras] {10.1093/mnras/stu2058}, \href
  {http://adsabs.harvard.edu/abs/2015MNRAS.446..521S} {446, 521}

\bibitem[\protect\citeauthoryear{{Schlafly} \& {Finkbeiner}}{{Schlafly} \&
  {Finkbeiner}}{2011}]{2011ApJ...737..103S}
{Schlafly} E.~F.,  {Finkbeiner} D.~P.,  2011, \mn@doi [\apj]
  {10.1088/0004-637X/737/2/103}, \href
  {http://adsabs.harvard.edu/abs/2011ApJ...737..103S} {737, 103}

\bibitem[\protect\citeauthoryear{{Scholtz} et~al.,}{{Scholtz}
  et~al.}{2018}]{2018MNRAS.475.1288S}
{Scholtz} J.,  et~al., 2018, \mn@doi [\mnras] {10.1093/mnras/stx3177}, \href
  {https://ui.adsabs.harvard.edu/abs/2018MNRAS.475.1288S} {475, 1288}

\bibitem[\protect\citeauthoryear{{Smith}, {Pacifici}, {Pasquali}  \&
  {Calder{\'o}n-Castillo}}{{Smith} et~al.}{2019}]{2019ApJ...876..145S}
{Smith} R.,  {Pacifici} C.,  {Pasquali} A.,   {Calder{\'o}n-Castillo} P.,
  2019, \mn@doi [\apj] {10.3847/1538-4357/ab1917}, \href
  {https://ui.adsabs.harvard.edu/abs/2019ApJ...876..145S} {876, 145}

\bibitem[\protect\citeauthoryear{{Springel}}{{Springel}}{2010}]{2010MNRAS.401..791S}
{Springel} V.,  2010, \mn@doi [\mnras] {10.1111/j.1365-2966.2009.15715.x},
  \href {http://adsabs.harvard.edu/abs/2010MNRAS.401..791S} {401, 791}

\bibitem[\protect\citeauthoryear{{Springel} \& {Hernquist}}{{Springel} \&
  {Hernquist}}{2003}]{2003MNRAS.339..312S}
{Springel} V.,  {Hernquist} L.,  2003, \mn@doi [\mnras]
  {10.1046/j.1365-8711.2003.06207.x}, \href
  {http://adsabs.harvard.edu/abs/2003MNRAS.339..312S} {339, 312}

\bibitem[\protect\citeauthoryear{{Springel}, {White}, {Tormen}  \&
  {Kauffmann}}{{Springel} et~al.}{2001}]{2001MNRAS.328..726S}
{Springel} V.,  {White} S. D.~M.,  {Tormen} G.,   {Kauffmann} G.,  2001,
  \mn@doi [\mnras] {10.1046/j.1365-8711.2001.04912.x}, \href
  {https://ui.adsabs.harvard.edu/abs/2001MNRAS.328..726S} {328, 726}

\bibitem[\protect\citeauthoryear{{Springel} et~al.,}{{Springel}
  et~al.}{2005}]{2005Natur.435..629S}
{Springel} V.,  et~al., 2005, \mn@doi [\nat] {10.1038/nature03597}, \href
  {http://adsabs.harvard.edu/abs/2005Natur.435..629S} {435, 629}

\bibitem[\protect\citeauthoryear{{Springel} et~al.,}{{Springel}
  et~al.}{2018}]{2018MNRAS.475..676S}
{Springel} V.,  et~al., 2018, \mn@doi [\mnras] {10.1093/mnras/stx3304}, \href
  {http://adsabs.harvard.edu/abs/2018MNRAS.475..676S} {475, 676}

\bibitem[\protect\citeauthoryear{{Suess}, {Kriek}, {Price}  \& {Barro}}{{Suess}
  et~al.}{2019}]{2019arXiv190410992S}
{Suess} K.~A.,  {Kriek} M.,  {Price} S.~H.,   {Barro} G.,  2019, arXiv
  e-prints, \href {https://ui.adsabs.harvard.edu/abs/2019arXiv190410992S} {p.
  arXiv:1904.10992}

\bibitem[\protect\citeauthoryear{{Thomas}, {Maraston}, {Bender}  \& {Mendes de
  Oliveira}}{{Thomas} et~al.}{2005}]{2005ApJ...621..673T}
{Thomas} D.,  {Maraston} C.,  {Bender} R.,   {Mendes de Oliveira} C.,  2005,
  \mn@doi [\apj] {10.1086/426932}, \href
  {http://adsabs.harvard.edu/abs/2005ApJ...621..673T} {621, 673}

\bibitem[\protect\citeauthoryear{{Thomas}, {Maraston}, {Schawinski}, {Sarzi}
  \& {Silk}}{{Thomas} et~al.}{2010}]{2010MNRAS.404.1775T}
{Thomas} D.,  {Maraston} C.,  {Schawinski} K.,  {Sarzi} M.,   {Silk} J.,  2010,
  \mn@doi [\mnras] {10.1111/j.1365-2966.2010.16427.x}, \href
  {http://adsabs.harvard.edu/abs/2010MNRAS.404.1775T} {404, 1775}

\bibitem[\protect\citeauthoryear{{Trayford} et~al.,}{{Trayford}
  et~al.}{2017}]{2017MNRAS.470..771T}
{Trayford} J.~W.,  et~al., 2017, \mn@doi [\mnras] {10.1093/mnras/stx1051},
  \href {http://adsabs.harvard.edu/abs/2017MNRAS.470..771T} {470, 771}

\bibitem[\protect\citeauthoryear{{Trujillo} et~al.,}{{Trujillo}
  et~al.}{2006}]{2006MNRAS.373L..36T}
{Trujillo} I.,  et~al., 2006, \mn@doi [\mnras]
  {10.1111/j.1745-3933.2006.00238.x}, \href
  {http://adsabs.harvard.edu/abs/2006MNRAS.373L..36T} {373, L36}

\bibitem[\protect\citeauthoryear{{Trussler}, {Maiolino}, {Maraston}, {Peng},
  {Thomas}, {Goddard}  \& {Lian}}{{Trussler}
  et~al.}{2020}]{2020arXiv200601154T}
{Trussler} J.,  {Maiolino} R.,  {Maraston} C.,  {Peng} Y.,  {Thomas} D.,
  {Goddard} D.,   {Lian} J.,  2020, arXiv e-prints, \href
  {https://ui.adsabs.harvard.edu/abs/2020arXiv200601154T} {p. arXiv:2006.01154}

\bibitem[\protect\citeauthoryear{{Vazdekis}, {S{\'a}nchez-Bl{\'a}zquez},
  {Falc{\'o}n-Barroso}, {Cenarro}, {Beasley}, {Cardiel}, {Gorgas}  \&
  {Peletier}}{{Vazdekis} et~al.}{2010}]{2010MNRAS.404.1639V}
{Vazdekis} A.,  {S{\'a}nchez-Bl{\'a}zquez} P.,  {Falc{\'o}n-Barroso} J.,
  {Cenarro} A.~J.,  {Beasley} M.~A.,  {Cardiel} N.,  {Gorgas} J.,   {Peletier}
  R.~F.,  2010, \mn@doi [\mnras] {10.1111/j.1365-2966.2010.16407.x}, \href
  {https://ui.adsabs.harvard.edu/abs/2010MNRAS.404.1639V} {404, 1639}

\bibitem[\protect\citeauthoryear{{Vogelsberger} et~al.,}{{Vogelsberger}
  et~al.}{2014a}]{2014MNRAS.444.1518V}
{Vogelsberger} M.,  et~al., 2014a, \mn@doi [\mnras] {10.1093/mnras/stu1536},
  \href {http://adsabs.harvard.edu/abs/2014MNRAS.444.1518V} {444, 1518}

\bibitem[\protect\citeauthoryear{{Vogelsberger} et~al.,}{{Vogelsberger}
  et~al.}{2014b}]{2014Natur.509..177V}
{Vogelsberger} M.,  et~al., 2014b, \mn@doi [\nat] {10.1038/nature13316}, \href
  {http://adsabs.harvard.edu/abs/2014Natur.509..177V} {509, 177}

\bibitem[\protect\citeauthoryear{{Weinberger} et~al.,}{{Weinberger}
  et~al.}{2017}]{2017MNRAS.465.3291W}
{Weinberger} R.,  et~al., 2017, \mn@doi [\mnras] {10.1093/mnras/stw2944}, \href
  {http://adsabs.harvard.edu/abs/2017MNRAS.465.3291W} {465, 3291}

\bibitem[\protect\citeauthoryear{{Weinmann}, {van den Bosch}, {Yang}  \&
  {Mo}}{{Weinmann} et~al.}{2006}]{2006MNRAS.366....2W}
{Weinmann} S.~M.,  {van den Bosch} F.~C.,  {Yang} X.,   {Mo} H.~J.,  2006,
  \mn@doi [\mnras] {10.1111/j.1365-2966.2005.09865.x}, \href
  {https://ui.adsabs.harvard.edu/abs/2006MNRAS.366....2W} {366, 2}

\bibitem[\protect\citeauthoryear{{Wetzel}, {Tinker}  \& {Conroy}}{{Wetzel}
  et~al.}{2012}]{2012MNRAS.424..232W}
{Wetzel} A.~R.,  {Tinker} J.~L.,   {Conroy} C.,  2012, \mn@doi [\mnras]
  {10.1111/j.1365-2966.2012.21188.x}, \href
  {https://ui.adsabs.harvard.edu/abs/2012MNRAS.424..232W} {424, 232}

\bibitem[\protect\citeauthoryear{{Wilkinson} et~al.,}{{Wilkinson}
  et~al.}{2015}]{2015MNRAS.449..328W}
{Wilkinson} D.~M.,  et~al., 2015, \mn@doi [\mnras] {10.1093/mnras/stv301},
  \href {https://ui.adsabs.harvard.edu/abs/2015MNRAS.449..328W} {449, 328}

\bibitem[\protect\citeauthoryear{{Yang}, {Mo}, {van den Bosch}, {Pasquali},
  {Li}  \& {Barden}}{{Yang} et~al.}{2007}]{2007ApJ...671..153Y}
{Yang} X.,  {Mo} H.~J.,  {van den Bosch} F.~C.,  {Pasquali} A.,  {Li} C.,
  {Barden} M.,  2007, \mn@doi [\apj] {10.1086/522027}, \href
  {http://adsabs.harvard.edu/abs/2007ApJ...671..153Y} {671, 153}

\bibitem[\protect\citeauthoryear{{Zibetti}, {Gallazzi}, {Hirschmann},
  {Consolandi}, {Falc{\'o}n-Barroso}, {van de Ven}  \& {Lyubenova}}{{Zibetti}
  et~al.}{2020}]{2020MNRAS.491.3562Z}
{Zibetti} S.,  {Gallazzi} A.~R.,  {Hirschmann} M.,  {Consolandi} G.,
  {Falc{\'o}n-Barroso} J.,  {van de Ven} G.,   {Lyubenova} M.,  2020, \mn@doi
  [\mnras] {10.1093/mnras/stz3205}, \href
  {https://ui.adsabs.harvard.edu/abs/2020MNRAS.491.3562Z} {491, 3562}

\bibitem[\protect\citeauthoryear{{da Cunha}, {Charlot}  \& {Elbaz}}{{da Cunha}
  et~al.}{2008}]{2008MNRAS.388.1595D}
{da Cunha} E.,  {Charlot} S.,   {Elbaz} D.,  2008, \mn@doi [\mnras]
  {10.1111/j.1365-2966.2008.13535.x}, \href
  {https://ui.adsabs.harvard.edu/abs/2008MNRAS.388.1595D} {388, 1595}

\bibitem[\protect\citeauthoryear{{van Dokkum} et~al.,}{{van Dokkum}
  et~al.}{2010}]{2010ApJ...709.1018V}
{van Dokkum} P.~G.,  et~al., 2010, \mn@doi [\apj]
  {10.1088/0004-637X/709/2/1018}, \href
  {http://adsabs.harvard.edu/abs/2010ApJ...709.1018V} {709, 1018}

\bibitem[\protect\citeauthoryear{{van den Bosch}, {Pasquali}, {Yang}, {Mo},
  {Weinmann}, {McIntosh}  \& {Aquino}}{{van den Bosch}
  et~al.}{2008}]{2008arXiv0805.0002V}
{van den Bosch} F.~C.,  {Pasquali} A.,  {Yang} X.,  {Mo} H.~J.,  {Weinmann} S.,
   {McIntosh} D.~H.,   {Aquino} D.,  2008, arXiv e-prints, \href
  {https://ui.adsabs.harvard.edu/abs/2008arXiv0805.0002V} {p. arXiv:0805.0002}

\bibitem[\protect\citeauthoryear{{van der Wel} et~al.,}{{van der Wel}
  et~al.}{2014}]{2014ApJ...788...28V}
{van der Wel} A.,  et~al., 2014, \mn@doi [\apj] {10.1088/0004-637X/788/1/28},
  \href {http://adsabs.harvard.edu/abs/2014ApJ...788...28V} {788, 28}

\makeatother
\end{thebibliography}

	
	
	\appendix


	\bsp	
	\label{lastpage}
\end{document}